\documentclass[12pt]{article}
\RequirePackage[OT1]{fontenc}

\usepackage{amsbsy,amsfonts,amsmath,amssymb,amsthm}
\usepackage{array}
\usepackage{booktabs}
\usepackage{color, dsfont}
\usepackage{enumerate}
\usepackage{epsf,epsfig}
\usepackage{fancyhdr}
\usepackage{fancyvrb}
\usepackage{float}
\usepackage{framed}
\usepackage{graphicx}
\usepackage{hyperref}  
\usepackage{lmodern}
\usepackage{ltablex} %long table to break into multiple pages with X column
\usepackage{makecell}
\usepackage{mathtools}
\usepackage{multirow}
\usepackage{orcidlink}
\usepackage{pdflscape}
\usepackage{psfrag}
\usepackage{ragged2e} 
\usepackage{rotating}
\usepackage[sectionbib]{natbib}
\usepackage{sectsty, secdot}
\usepackage{seqsplit}
\usepackage{seqsplit}
\usepackage{subcaption}
\usepackage{tabularx}
\usepackage{threeparttable}
\usepackage{thumbpdf}
\usepackage{titlesec} % For section title formatting
\usepackage{url} 
\usepackage{verbatim}

\newcolumntype{L}{>{\RaggedRight\arraybackslash}X}  % 
\newcolumntype{C}{>{\Centering\arraybackslash}X}    % 
\newcolumntype{R}{>{\RaggedLeft\arraybackslash}X}   %

\def\thefigure{\arabic{figure}}
\def\thetable{\arabic{table}}
\begin{document}

\title{\Large\bf ForLion: An R Package for Finding Optimal Experimental Designs with Mixed Factors}
\author{Siting Lin\footnote{These authors contributed equally to this work.} $^1$, Yifei Huang$^*$$^2$, and Jie Yang$^1$\\ 
	$^1$University of Illinois at Chicago and\\ $^2$Astellas Pharma Global Development, Inc.
}

\maketitle

\begin{abstract}
Optimal design is crucial for experimenters to maximize the information collected from experiments and estimate the model parameters most accurately. ForLion algorithms have been proposed to find D-optimal designs for experiments with mixed types of factors. In this paper, we introduce the ForLion package which implements the ForLion algorithm to construct locally D-optimal designs and the Expected Weighted (EW) ForLion algorithm to generate robust EW D-optimal designs, which maximize the determinant of the expected Fisher information matrix under parameter uncertainty. The package supports experiments under linear models (LM), generalized linear models (GLM), and multinomial logistic models (MLM) with continuous, discrete, or mixed-type factors. It provides both optimal approximate designs and an efficient function converting approximate designs into exact designs with integer-valued allocations of experimental units. Tutorials are included to show the package's usage across different scenarios.
\end{abstract}

\section{Introduction}\label{sec:introduction}

The study of optimal designs can be traced back to \cite{smith1918standard} on regression problems for univariate polynomials of order up to six \citep{fedorov2014}. Later in the 20th century, the optimal design theory has been expanded to encompass various statistical models and optimality criteria \citep{fedorov1972, silvey1980, pukelsheim1993, atkinson2007, fedorov2014}. 
Despite significant advancements on linear regression models, generalized linear models (GLM) for more general univariate responses \citep{pmcc1989, dobson2018, khuri2006, stufken2012} and multinomial logistic models (MLM) for categorical responses \citep{pmcc1995, atkinson1999, bu2020} have been widely used in practice, but are much more difficult in the optimal design theory, because their Fisher information matrices depend on model parameters.

Package {\bf AlgDesign} (version 1.2.1.2, see \cite{wheeler2025package}) offers tools to construct optimal designs with exact and approximate allocations, focusing on linear models including polynomial forms. It allows mixed factors by generating a finite candidate list of experimental settings and employs the Fedorov's exchange algorithm \citep{fedorov1972} for optimization under D-, A-, and I-criteria. Package {\bf OptimalDesign} (version 1.0.2.1, see \cite{harman2025package, harman2020randomized}) finds D-, A-, I-, and c-efficient designs for linear models (LM), generalized linear models (GLM), some dose-response, and certain survival models, with mixed factors by discretizing the continuous factors first. \cite{harman2021optimal} further proposed a grid-exploration method for mixed-factor D-optimal designs on cuboid grids under linear, generalized linear, and nonlinear regression models. Package {\bf ICAOD} (version 1.0.1, see \cite{masoudi2022package}) provides tools for finding locally, minimax, and Bayesian D-optimal designs, as well as user-specified optimality criteria, for nonlinear statistical models. It implements the imperialist competitive algorithm \citep{masoudi2022icaod} for designs involving continuous factors only.
Package {\bf PFIM} (version 6.1, see \cite{mentre2024package, dumont2018pfim}) computes D-optimal designs for nonlinear mixed-effects models (NLMEM), which is particularly useful for pharmacokinetic/pharmacodynamic models (PKPD). It handles designs either with discrete factors only or with continuous factors only. 
Package {\bf idefix} (version 1.1.0, see \cite{traets2025package, traets2020generating}) is designed to generate D-efficient and Bayesian D-efficient optimal designs for multinomial logit (MNL) and mixed logit (MIXL) models for discrete choice experiments (DCE). However, these existing packages are designed for experiments with discrete factors only, continuous factors only, or mixed factors by discretizing the continuous factors first. There has been limited progress in constructing efficient designs that incorporate both discrete and continuous factors \citep{huang2024forlion}. 

The {\bf ForLion} package,  available at the Comprehensive R Archive Network (CRAN, \url{ https://CRAN.R-project.org/package=ForLion}), complements existing software by providing computational tools for constructing D-optimal experimental designs involving both types of factors under fairly general parametric statistical models using the ForLion algorithm proposed by \cite{huang2024forlion}. A key contribution of the package is to provide practical computational tools for constructing D-optimal designs for experiments involving multinomial or ordinal qualitative responses under the MLM framework, while using a unified strategy for mixed-factor design problems. In addition, for GLMs and some MLMs, the package adopts internal optimizations via analytic solutions, which can significantly improve the computational efficiency \citep{huang2024forlion}. Different from discretizing the continuous factors first, it starts from a randomly generated or user-provided initial design and iteratively applies merging, lift-one, and deletion steps to control the number of support points. In the new-point step, it enumerates the combinations of discrete-factor levels and optimizes continuous-factor levels conditional on each discrete-factor setting. As a result, it tends to reduce the number of distinct experimental settings while maintaining high efficiency of designs. It covers linear models (LM), generalized linear models (GLM), and general multinomial logistic models (MLM). Note that the MLM here is much broader than the MNL for discrete choice experiments. It includes three more classes of models for ordinal responses, namely cumulative, adjacent-categories, and continuation-ratio logit models \citep{bu2020}. Furthermore, to overcome the issue caused by the dependence of Fisher information matrices on model parameters, we also implement the EW ForLion algorithm proposed by \cite{lin2025ew} for constructing robust optimal designs against unknown model parameters for LM, GLM, and MLM as well. Here, the EW criterion is a practical surrogate to Bayesian D-optimality under parameter uncertainty, and the resulting computation can be carried out within a similar computational framework to ForLion. The motivation and theoretical justifications of the EW criterion for mixed-factor design problems can be found in \cite{lin2025ew}.

In Section~\ref{sec:ForLion}, we outline the theoretical foundations of the {\bf ForLion} package, including the ForLion algorithm, the lift-one algorithm, the EW ForLion algorithm, and a rounding algorithm. In Section~\ref{sec:structure}, we introduce the structure and function arguments of the {\bf ForLion} package, followed by illustrative examples of the package applications, as well as the interpretations of the key results in Section~\ref{sec:example}. We summarize and conclude in Section~\ref{sec:summary}. 

\section{Method}\label{sec:ForLion}

We consider a mixed-factors experiment under a general statistical model $M(\mathbf x; \boldsymbol \theta)$ with parameter(s) $\boldsymbol \theta \in \boldsymbol{\Theta} \subseteq \mathbb{R}^p$ and experimental setting ${\mathbf x} \in {\cal X} \subset \mathbb{R}^d$, where $\boldsymbol{\Theta}$ is called the parameter space, and ${\cal X}$ is called the design region or design space. For typical applications, ${\cal X}$ is compact. Among the $d$ factors, without any loss of generality, we assume that the first $k$ factors are continuous and the last $d-k$ are discrete, for $0\le k \le d$. Following \cite{lin2025ew}, in package {\bf ForLion}, we cover three scenarios: {\it (i)} if $k=0$, ${\cal X} = {\cal D}$ consists of a predetermined finite list of level combinations of discrete factors; {\it (ii)} if $1\leq k\leq d-1$, ${\cal X} = \prod_{j=1}^k I_j\times {\cal D}$, with $I_j=[a_j, b_j]$ being a finite closed interval; and {\it (iii)} if $k=d$, ${\cal X} = \prod_{j=1}^d I_j$~.

An experimental design $\boldsymbol{\xi}$ considered in this paper consists of $m$ distinct design points, $\mathbf x_1, \dots, \mathbf x_m$ $\in {\cal X}$, and real-valued proportions ${\mathbf w} = (w_1, \ldots, w_m)^\top$ satisfying $w_i\geq 0$ for each $i$ and  $\sum_{i=1}^m w_i = 1$, known as an approximate allocation of the experimental units. In practice, we also look for integer-valued assignments ${\mathbf n} = (n_1, \ldots, n_m)^\top$ given $N=\sum_{i=1}^m n_i$~, known as an exact allocation. The collection of approximate designs is denoted by $\boldsymbol{\Xi} = \{\boldsymbol{\xi} = \{({\mathbf x}_i, w_i), i=1, \ldots, m\} \mid m\geq 1, {\mathbf x}_i \in {\cal X}, w_i\geq 0, \sum_{i=1}^m w_i = 1\}$.
Under regularity conditions, the Fisher information matrix associated with the design $\boldsymbol{\xi}$ can be denoted as ${\mathbf F}(\boldsymbol{\xi}, \boldsymbol{\theta}) = \sum_{i=1}^m w_i {\mathbf F}({\mathbf x}_i, \boldsymbol{\theta}) \in \mathbb{R}^{p\times p}$, where ${\mathbf F}({\mathbf x}_i, \boldsymbol{\theta})$ is the Fisher information associated with ${\mathbf x}_i$~.

When the experimenter has a good idea about the values of $\boldsymbol \theta$, following \cite{huang2024forlion}, we look for a locally D-optimal design that maximizes $f_{\boldsymbol{\theta}}(\boldsymbol \xi)=| \mathbf F(\boldsymbol \xi, \boldsymbol \theta)|$ by implementing the ForLion algorithm (see Section~\ref{sec:ForLion algorithms}). 
In practice, however, the experimenter may not be certain about the true parameter values. In package {\bf ForLion}, we offer two options for the users. With a prespecified prior distribution or probability measure $Q(\cdot)$ on $\boldsymbol{\Theta}$, we look for an EW D-optimal design \citep{atkinson2007, ymm2016, ytm2016, bu2020, huang2025constrained, lin2025ew} that maximizes 
\begin{equation}\label{eq:f_EW}
f_{\rm EW}(\boldsymbol{\xi}) = |E\{{\mathbf F}(\boldsymbol{\xi}, \boldsymbol{\Theta})\}| = \left|\int_{\boldsymbol{\Theta}} {\mathbf F}(\boldsymbol{\xi}, \boldsymbol{\theta}) Q(d\boldsymbol{\theta})\right|\ 
\end{equation}
by implementing the EW ForLion algorithm proposed by \cite{lin2025ew}, also called an integral-based EW D-optimal design. When a dataset from a pilot study is available, or if the integral in \eqref{eq:f_EW} is difficult to calculate, we look for a sample-based EW D-optimal design \citep{lin2025ew} that maximizes
\begin{equation}\label{eq:f_SEW}
f_{\rm SEW}(\boldsymbol{\xi}) = |\hat{E}\{{\mathbf F}(\boldsymbol{\xi}, \boldsymbol{\Theta})\}| = \left|\frac{1}{B}\sum_{j=1}^B {\mathbf F}(\boldsymbol{\xi}, \hat{\boldsymbol{\theta}}_j) \right|\ ,
\end{equation}
where $\{\hat{\boldsymbol{\theta}}_1, \ldots, \hat{\boldsymbol{\theta}}_B\}$ are either estimated from bootstrapped samples from the pilot dataset, or sampled from the prior distribution $Q(\cdot)$ on $\boldsymbol{\Theta}$ (see Section~\ref{sec:EW_ForLion}). 

To compare two designs, we report the relative efficiency of design $\boldsymbol{\xi}_1$ with respect to design $\boldsymbol{\xi}_2$ in terms of their criterion values. For local D-optimality, we define
\[
\mathrm{Eff}_{\mathrm{loc}}(\boldsymbol{\xi}_1;\boldsymbol{\xi}_2)=\left(\frac{|{\mathbf F}(\boldsymbol{\xi}_{\rm 1}, \boldsymbol\theta)|}{|{\mathbf F}(\boldsymbol{\xi}_{\rm 2}, \boldsymbol\theta)|}\right)^{1/p}\ ,
\]
where $p$ is the number of parameters.
For EW D-optimality, the relative efficiency is defined by
\[
\begin{array}{c@{\qquad\text{or}\qquad}c}
\displaystyle
\mathrm{Eff}_{\mathrm{EW}}(\boldsymbol{\xi}_1;\boldsymbol{\xi}_2)
=\left(\frac{|E\{{\mathbf F}({\boldsymbol{\xi}}_{1}, \boldsymbol{\Theta})\}|}
{|E\{{\mathbf F}({\boldsymbol{\xi}}_{2}, \boldsymbol{\Theta})\}|}\right)^{1/p}
&
\displaystyle
\mathrm{Eff}_{\mathrm{SEW}}(\boldsymbol{\xi}_1;\boldsymbol{\xi}_2)
=\left(\frac{|\hat{E}\{{\mathbf F}({\boldsymbol{\xi}}_{1}, \boldsymbol{\Theta})\}|}
{|\hat{E}\{{\mathbf F}({\boldsymbol{\xi}}_{2}, \boldsymbol{\Theta})\}|}\right)^{1/p}
\end{array}\ .
\]

According to Theorem~1 of \cite{huang2024forlion}, unlike stochastic optimization algorithms such as particle swarm optimization (PSO), the designs found by the ForLion algorithm are locally D-optimal when the converging condition based on the maximized sensitivity function is met. Based on Corollary~1 of \cite{lin2025ew}, the designs constructed by the EW ForLion algorithm are EW D-optimal under a similar condition, which are not only robust against unknown parameters, but also computationally much easier to find than traditional Bayesian D-optimal designs, while keeping high-efficiency \citep{ymm2016, ytm2016, bu2020, lin2025ew}. Nevertheless, from a practical point of view, such an algorithm may stop before reaching a D-optimal design, if it fails to find the global maxima for the sensitivity function (see Remark~4 of \cite{huang2024forlion}). As a common practice, one may explore multiple random starting points during the maximization to increase the chance of success.

\subsection{ForLion algorithm} \label{sec:ForLion algorithms}

\begin{figure}[ht]
     \centering
     \includegraphics[width=1\textwidth]{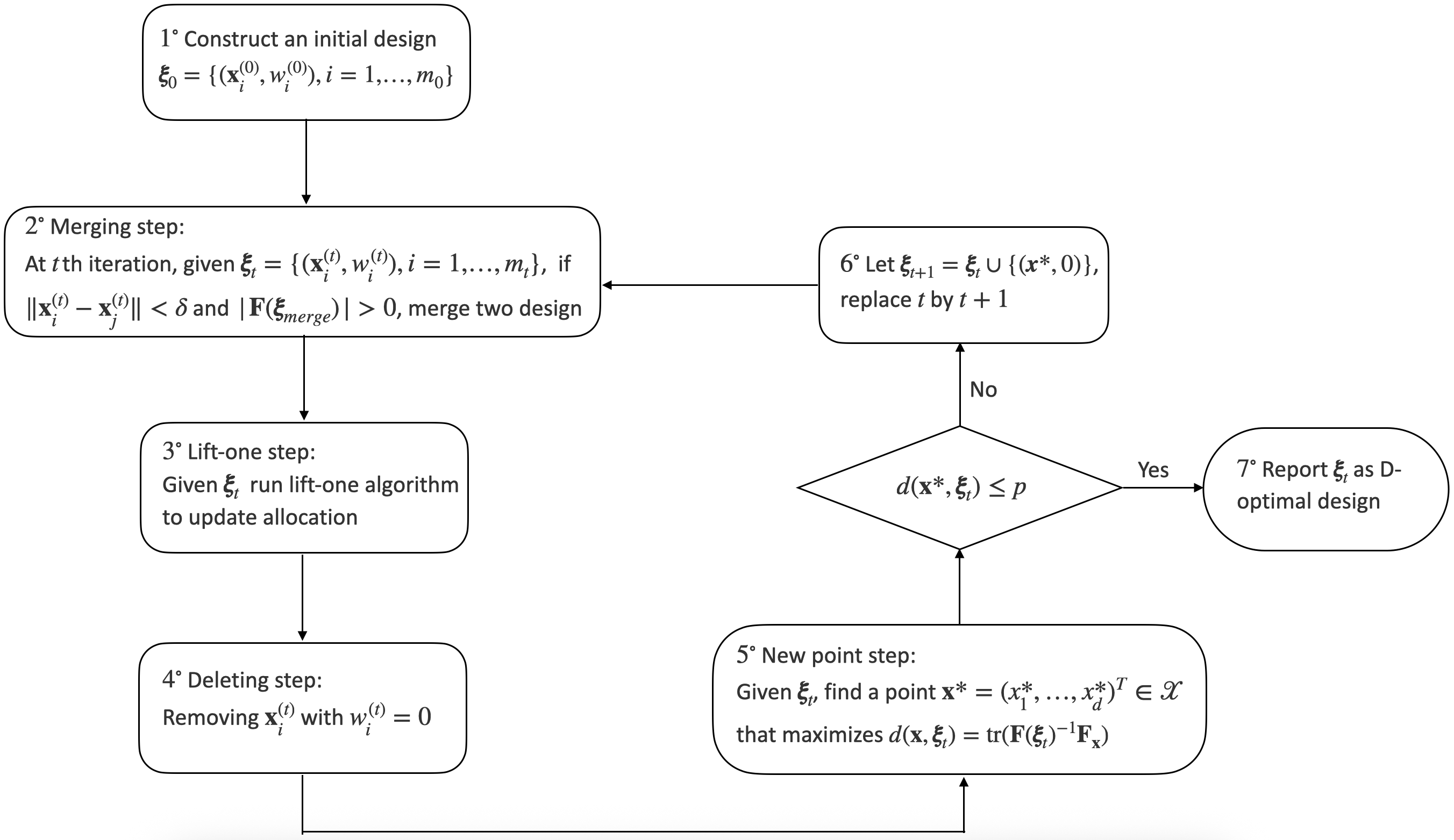}
     \caption{An outline of the ForLion algorithm under a general statistical model.}\label{fig:ForLion_general}
 \end{figure}

In this section, we assume that $\boldsymbol{\theta} \in \boldsymbol{\Theta}$ is known. To simplify notations, we let ${\mathbf F}_{\mathbf x}$ and ${\mathbf F}(\boldsymbol{\xi})$ stand for ${\mathbf F}({\mathbf x}, \boldsymbol{\theta})$ and ${\mathbf F}(\boldsymbol{\xi}, \boldsymbol{\theta})$, respectively. The ForLion algorithm looks for a locally D-optimal design $\boldsymbol{\xi}_* \in \boldsymbol{\Xi}$ that maximizes $|{\mathbf F}(\boldsymbol\xi)|$.  It begins with an initial design $\boldsymbol{\xi}_0 \in \boldsymbol{\Xi}$ in Step $1^\circ$ (see Figure~\ref{fig:ForLion_general}) satisfying $|{\mathbf{F}}({\boldsymbol{\xi}}_0)| > 0$, with a prespecified distance threshold $\delta_{0} > 0$ between any two design points, such that, $\|{\mathbf{x}}_i^{(0)} - {\mathbf{x}}_j^{(0)}\| \geq \delta_{0}$ for any $i \neq j$. To reduce the number of distinct experimental settings, which in practice often indicates reduced experimental time and cost, at the beginning of each round of iterations, the ForLion algorithm reduces the number of design points by merging those points in close distance (i.e., less than $\delta$), which is performed only if the resulting design still satisfies $|{\mathbf{F}}({\boldsymbol{\xi}}_{\rm merge})|>0$ (see Step~$2^\circ$ in Figure~\ref{fig:ForLion_general}). According to Figure~S2 in the Supplementary Material of \cite{huang2024forlion}, larger $\delta$ may yield fewer support points and increased minimum distance among support points. On the other hand, however, too large $\delta$ may lead to unnecessary merges and possible loss in D-efficiency. In practice, one may choose an appropriate $\delta$ to balance the number of design points and D-efficiency. In Step~$3^\circ$, the lift-one algorithm \citep{ym2015, ymm2016, huang2025constrained} is employed to update the approximate allocation for the current design $\boldsymbol{\xi}_t$~. In Step~$4^\circ$, the design points with zero allocation are removed from the design.

In Step~$5^\circ$, the ForLion algorithm identifies a new design point $\mathbf{x}^*$ that maximizes the sensitivity function $d(\mathbf{x}, \boldsymbol{\xi}_t) = \operatorname{tr}(\mathbf{F}(\boldsymbol{\xi}_t)^{-1} \mathbf{F}_{\mathbf{x}})$. When both discrete and continuous factors are present, we enumerate the finite combinations of discrete-factor levels $\mathbf{x}_{(2)}$~. For each possible $\mathbf{x}_{(2)}$~, we maximize $d\bigl((\mathbf{x}_{(1)},\mathbf{x}_{(2)}), \boldsymbol{\xi}_t\bigr)$ over the continuous-factor levels $\mathbf{x}_{(1)}$ using the L-BFGS-B quasi-Newton method, and then select the $\mathbf{x}_{(2)}$ that attains the largest optimized sensitivity function value. According to Theorem 2.2 of \cite{fedorov2014} and Theorem 1 in \cite{huang2024forlion}, if $d(\mathbf{x}^*, \boldsymbol{\xi}_t) \leq p$, then $\boldsymbol{\xi}_t$ is D-optimal; otherwise, $\mathbf{x}^*$ is added to $\boldsymbol{\xi}_t$ with an initial zero allocation in Step~$6^\circ$, and the process returns to Step~$2^\circ$. In this package, we relax the stopping rule using a prespecified level of relative tolerance \texttt{rel.tol}. The algorithm stops when $d(\mathbf{x}^*, \boldsymbol{\xi}_t) \leq (1+\texttt{rel.tol})p$ and reports the resulting design as a numerically D-optimal design up to the relative tolerance. The outline of the ForLion algorithm is provided in Figure~\ref{fig:ForLion_general}, with detailed descriptions available in Algorithm~1 of \cite{huang2024forlion}.

\begin{figure}[htbp]
     \centering
     \includegraphics[width=1\textwidth]{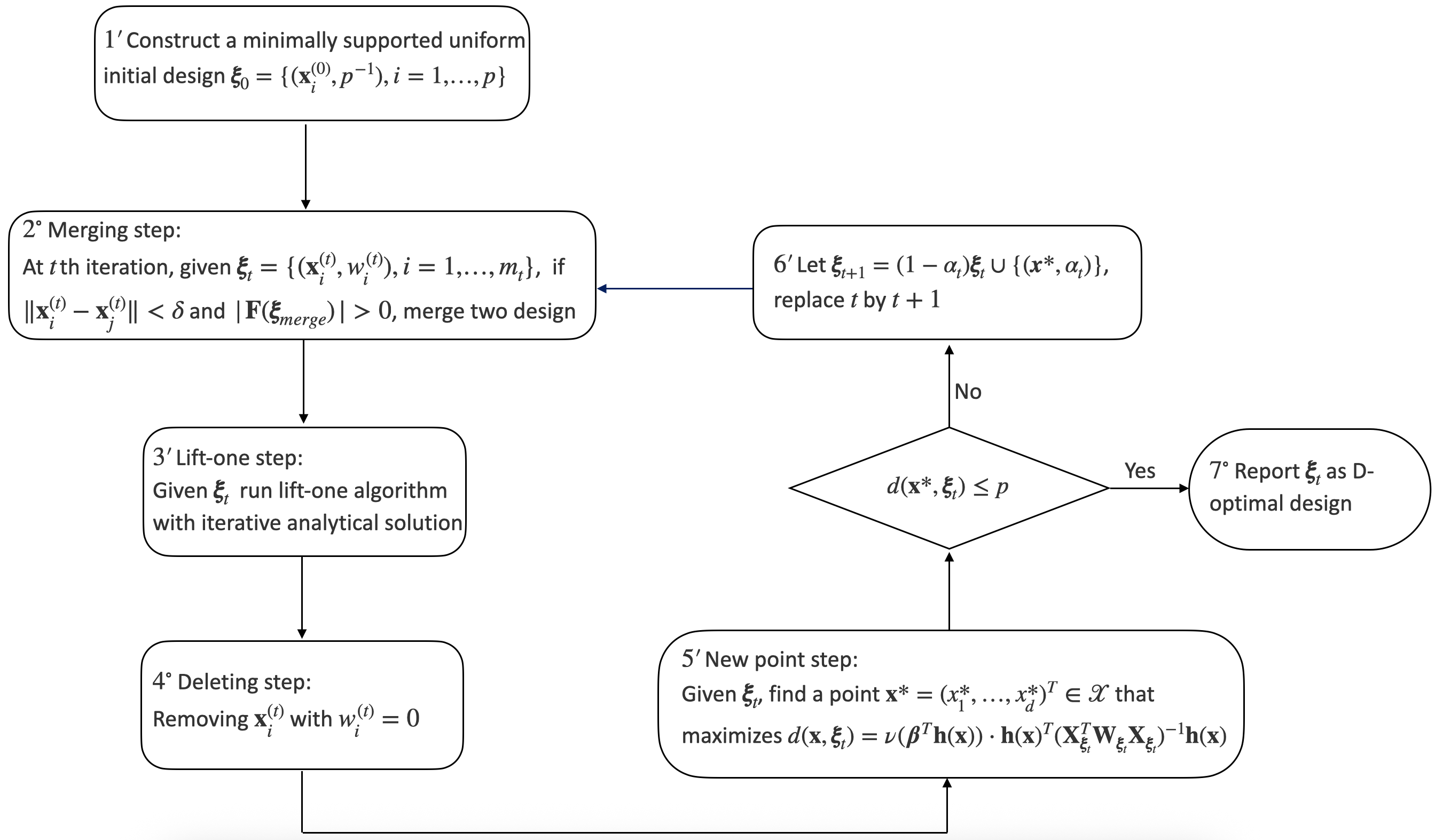}
     \caption{An outline of the GLM-adapted ForLion algorithm.}
     \label{fig:ForLion_GLM}
 \end{figure}

To speed up the ForLion algorithm for LM and GLM, we replace Steps $1^{\circ}$, $3^{\circ}$, $5^{\circ}$, and $6^{\circ}$ in Figure~\ref{fig:ForLion_general} with Steps $1^{\prime}$, $3^{\prime}$, $5^{\prime}$, and $6^{\prime}$ in Figure~\ref{fig:ForLion_GLM}, respectively, and implement the analytical solutions for GLM derived by \cite{ym2015} and \cite{huang2024forlion}. More specifically, in Step~$1^{\prime}$, we replace the random initial design with a minimally supported uniform design. In Step~$3^{\prime}$, we adopt the analytic solutions provided by \cite{ym2015} for the lift-one algorithm (see Section~\ref{sec:liftone_algorithm}) under a GLM. In Step~$5^{\prime}$, we adopt the simplified form of the sensitivity function as described in Theorem 4 of \cite{huang2024forlion}. In Step~$6^{\prime}$, we assign an initial weight $\alpha_t$ instead of zero to the new design point $\mathbf{x}^*$, based on Theorem~5 of \cite{huang2024forlion}. Those modifications accelerate the ForLion algorithm for LM and GLM. The specialized ForLion algorithm is illustrated by Figure~\ref{fig:ForLion_GLM}, with further details provided in Section~4 of \cite{huang2024forlion}.

\subsection{Lift-one algorithm}
\label{sec:liftone_algorithm}

The lift-one algorithm implemented in Step~$3^{\circ}$ of the ForLion algorithm was originally proposed by \cite{ymm2016} and \cite{ym2015} for GLMs, and then extended for cumulative link models by \cite{ytm2016}, and MLM by \cite{bu2020}. A general version and a constrained version of it can be found in \cite{huang2025constrained} and their Supplementary Material. It is highly efficient for finding the D-optimal approximate allocation ${\mathbf w} = (w_1, \ldots, w_m)^\top$ for a given set of design points $\{{\mathbf x}_1, \ldots, {\mathbf x}_m\}$. By adjusting the $i$th weight $w_i < 1$ while rescaling the remaining weights proportionally, the adjusted allocation is given by
\[
  \mathbf w_i(z) = \left( \frac{1-z}{1-w_i}w_1, \dots, \frac{1-z}{1-w_i}w_{i-1}, z, \frac{1-z}{1-w_i}w_{i+1}, \dots, \frac{1-z}{1-w_i}w_m \right)^\top
\]
with $z\in [0,1]$, which converts a multi-dimensional optimization problem to a one-dimensional optimization problem. The lift-one algorithm is closely related to the vertex direction method (VDM) \citep{wynn1970, fedorov1972, yu2011d, fedorov2025model}, which, however, chooses the design point that maximizes the directional derivative and keeps the updated weights positive. Unlike VDM, by picking up the design points in turn, the optimal weight $z_*$ in the lift-one algorithm can be exactly zero, allowing the ForLion algorithm to reduce the number of support points and keep the intermediate designs sparse, which can improve its computational efficiency. This weight-nullifying property can also be found in the vertex exchange method (VEM) of \cite{bohning1986vertex} and the randomized exchange algorithm (REX) of \cite{harman2020randomized}. For GLMs and MLMs with five or less categories, in this package we implement analytic solutions for the optimal $z_* \in [0,1]$, which further improves the computational efficiency \citep{ym2015, lin2025ew}. By adjusting ${\mathbf w}_i(z)$ for each $i=1, \dots, m$ in a random order, the converged allocation is guaranteed to be D-optimal.

The lift-one algorithm has been shown to be computationally efficient in the specific settings considered in \cite{ymm2016}. In particular, the numerical comparisons in \cite{ymm2016} imply favorable computational performance of the lift-one algorithm relative to several commonly used optimization techniques, including Nelder-Mead, quasi-Newton, and simulated annealing, as well as popular design algorithms such as Fedorov-Wynn, multiplicative, and cocktail algorithms (see \cite{huang2024forlion} for a good review). It often achieves a design with a reduced number of support points, that is, the design points with positive weights.

\subsection{EW ForLion algorithm} \label{sec:EW_ForLion}

In this section, the model parameter vector $\boldsymbol{\theta} \in \boldsymbol{\Theta}$ is assumed to be unknown. Instead, either a prior distribution $Q(\cdot)$ on $\boldsymbol{\Theta}$ or a dataset obtained from a previous study is available.

Given a prior distribution or probability measure $Q(\cdot)$ on $\boldsymbol{\Theta}$, we adopt the EW ForLion algorithm proposed by \cite{lin2025ew} to find an integral-based EW D-optimal design $\boldsymbol{\xi}_*\in \boldsymbol{\Xi}$ that maximizes $f_{\rm EW}(\boldsymbol{\xi}) = |\int_{\boldsymbol{\Theta}} {\mathbf F}(\boldsymbol{\xi}, \boldsymbol{\theta}) Q(d\boldsymbol{\theta})|$ as defined in \eqref{eq:f_EW}.
Different from the ForLion algorithm maximizing $f_{\boldsymbol{\theta}}(\boldsymbol \xi)=| \mathbf F(\boldsymbol \xi, \boldsymbol \theta)|$ with a prespecified $\boldsymbol{\theta}$, the EW ForLion algorithm 
targets the expectation of the Fisher information matrix, $E\left\{ {\mathbf F}(\boldsymbol{\xi}, \boldsymbol{\Theta})\right\}$.
By computing the entry-wise expectation with respect to the prior distribution $Q(\cdot)$ on $\boldsymbol{\Theta}$, we obtain a $p\times p$ matrix as well. 
Commonly used prior distributions include uniform priors on bounded intervals, normal priors on $\mathbb{R}$, and Gamma priors on $(0, \infty)$ (see, e.g., \cite{huang2025constrained}). In this package, we calculate the corresponding integrals by using the \texttt{hcubature()} function in package {\bf cubature} \citep{balasubramanianpackage}, which applies an adaptive multidimensional integration method by subdividing hyper-rectangular domains.

Alternatively, if a dataset from a previous or pilot study is available, we may bootstrap it for $B$ times (e.g., $B=1000$). For the $j$th bootstrapped dataset, we fit the statistical model and obtain the parameter estimates   $\hat{\boldsymbol{\theta}}_j$~, for $j=1, \ldots, B$. In this package, we implement the EW ForLion algorithm to find a sample-based EW D-optimal design $\boldsymbol{\xi}_*$~, which maximizes $f_{\rm SEW}(\boldsymbol{\xi}) = |B^{-1} \sum_{j=1}^B {\mathbf F}(\boldsymbol{\xi}, \hat{\boldsymbol{\theta}}_j)|$ as defined in \eqref{eq:f_SEW}.
In practice, if the expected Fisher information matrix is difficult to calculate with respect to $Q(\cdot)$, e.g., for some MLM \citep{lin2025ew}, we may also simulate $\hat{\boldsymbol{\theta}}_1, \ldots, \hat{\boldsymbol{\theta}}_B$ from $Q(\cdot)$, and look for a sample-based EW D-optimal design. According to \cite{lin2025ew}, the resulting designs are fairly robust against different sets of parameter vectors in terms of relative efficiency.

When there is no confusion, we let ${\mathbf F}_{\mathbf x}$ represent $\int_{\boldsymbol{\Theta}} {\mathbf F}({\mathbf x}, \boldsymbol{\theta}) Q(d\boldsymbol{\theta})$ for integral-based EW optimality, or $B^{-1} \sum_{j=1}^B {\mathbf F}({\mathbf x}, \hat{\boldsymbol{\theta}}_j)$ for sample-based EW optimality. Similarly, we let ${\mathbf F}(\boldsymbol{\xi})$ represent $E\{{\mathbf F}(\boldsymbol{\xi}, \boldsymbol{\Theta})\}$ for integral-based EW optimality, or $\hat{E}\{{\mathbf F}(\boldsymbol{\xi}, \boldsymbol{\Theta})\}$ for sample-based EW optimality. Then Figure~\ref{fig:ForLion_general} may also be used for illustrating the EW ForLion algorithm, with more detailed descriptions in Algorithm~1 of \cite{lin2025ew}.

\subsection{Rounding algorithm}
\label{sec:rounding_algorithm}

Both the ForLion and EW ForLion algorithms are able to find optimal experimental settings in a continuous or mixed design region. In practice, the suggested experimental settings may need to be rounded up due to the sensitivity level of the experimental device or environmental control. For example, 149.2116 Gy as the gamma radiation level may need to be rounded up to 149.2 Gy due to the sensitivity level of the radiation device (see the emergence of house flies example in Section~\ref{sec:fly_example}). On the other hand, the obtained optimal approximate design may also need to be converted to an exact design given a total number $N$ of experimental units.

\begin{figure}[htbp]
 \centering
 \includegraphics[width=0.8\textwidth]{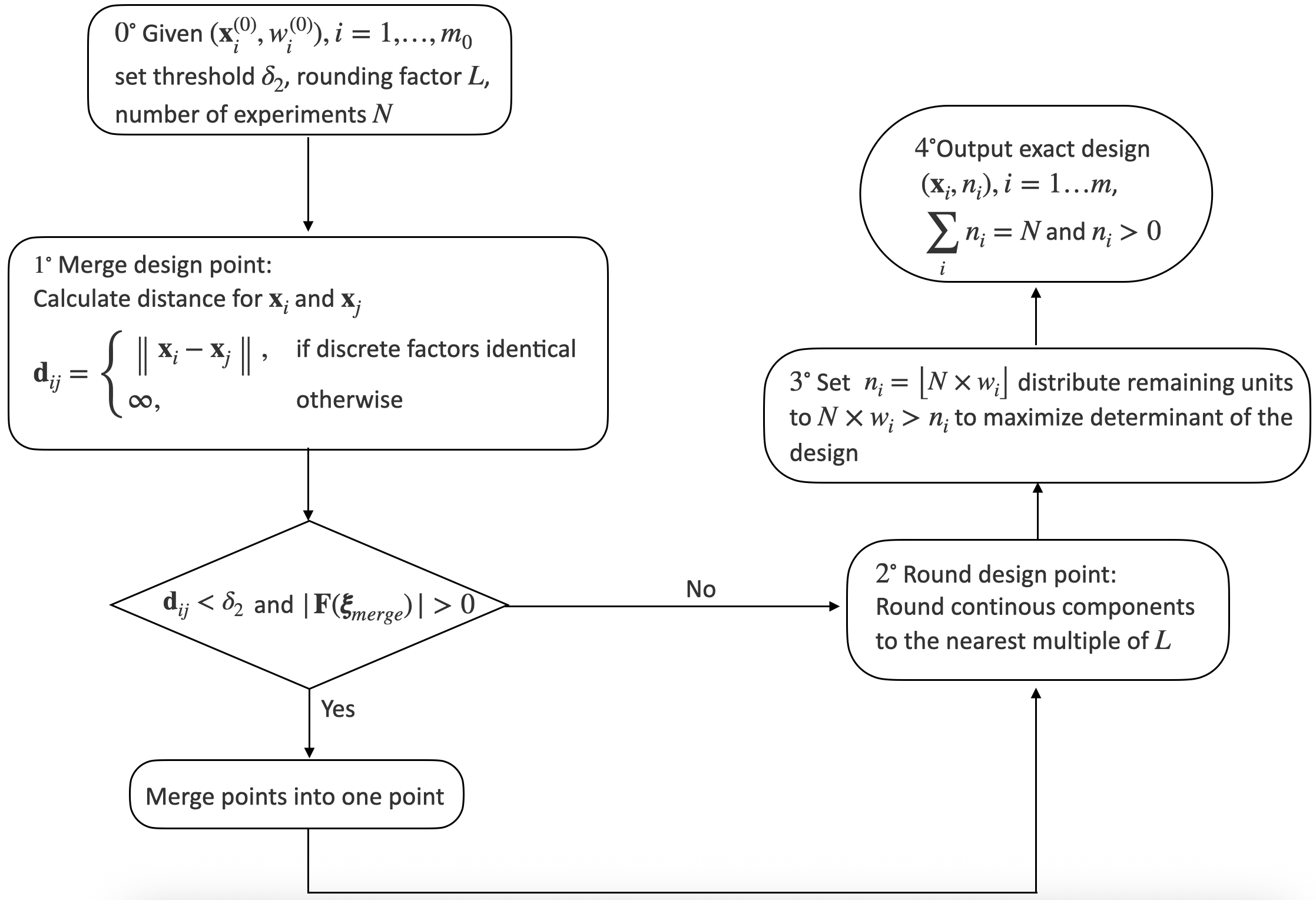}
\caption{An outline of the rounding algorithm.}
\label{fig:rounding_algo}
\end{figure}

In this package, we adopt the rounding algorithm proposed by \cite{lin2025ew} to convert an approximate design obtained by the ForLion or EW ForLion algorithm on a continuous or mixed region to an exact design with user-specified levels of grid points and $N$. 
It starts by merging design points based on a specified distance measure and a merging threshold $\delta_2$ (see Step~$1^\circ$ in Figure~\ref{fig:rounding_algo}). Then the levels of the continuous factors are rounded to the nearest multiples of user-defined grid levels (see Step~$2^\circ$). As for the approximate allocation $w_i$~, the rounding algorithm initializes the corresponding integer allocation $n_i = \lfloor{N \times w_i}\rfloor$, the largest integer no more than $Nw_i$~, and then allocates the remaining experimental units one by one to maximize the objective function (see Step~$3^\circ$). The resulting exact design is reported in Step~$4^\circ$. The outline of the rounding algorithm is displayed in Figure~\ref{fig:rounding_algo}, with more details provided in Algorithm~2 of \cite{lin2025ew} . 

Compared with optimal designs constructed directly on the same set of grid points, the design rounded from a ForLion design costs much less time, contains less distinct experimental settings, and maintains a high relative efficiency with respect to the ForLion design \citep{lin2025ew}.

\section{ForLion Package Structure}\label{sec:structure}

\begin{figure}[htb]
\centering
    \includegraphics[width=0.8\linewidth]{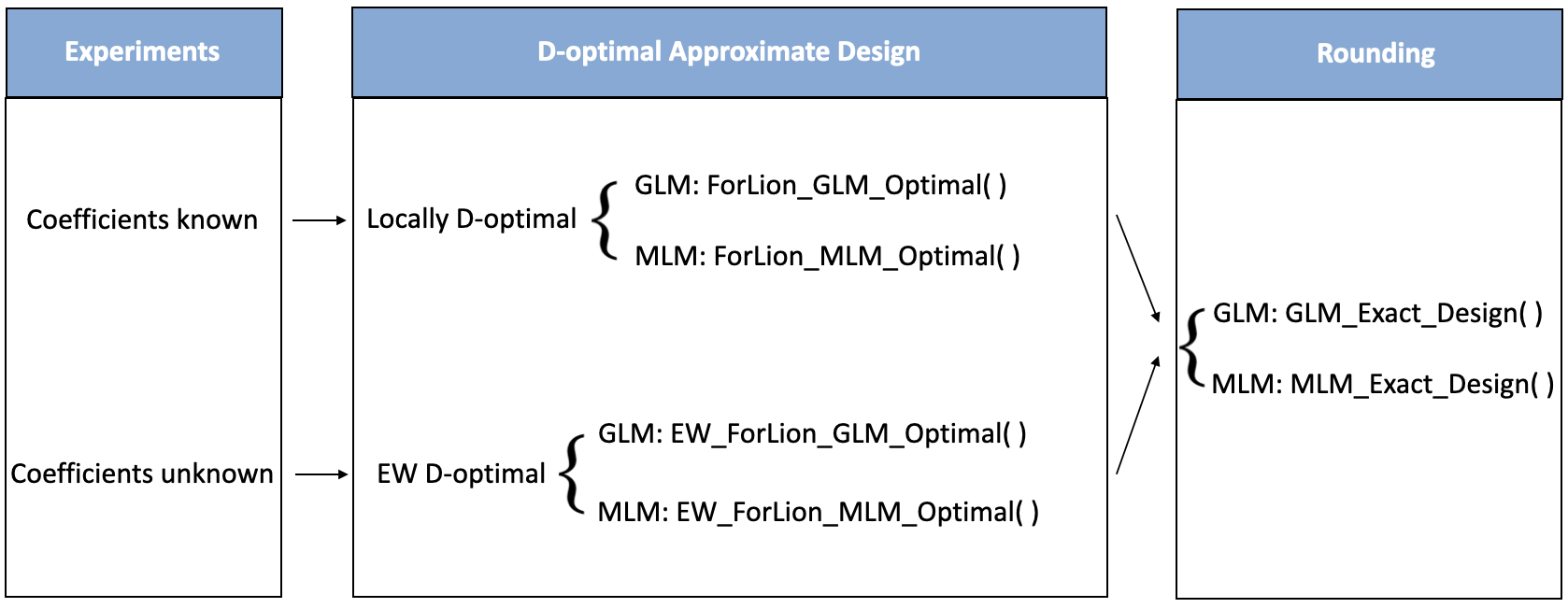}
    \caption{The key functions in the {\bf ForLion} package and the corresponding structure.}
    \label{fig:structure}
\end{figure}

The current version (0.4.0) of the {\bf ForLion} package \citep{huanglinpackage} supports finding D-optimal designs for experiments under parametric models with discrete factors only, continuous factors only, or mixed factors, including linear models (LM or GLM with ``identity'' link), logistic models (GLM with ``logit'' link) and other GLMs for binary responses (GLM with ``probit'', ``cloglog'', ``loglog'', and ``cauchit'' links), Poisson models (GLM with ``log'' link), baseline-category logit models or multiclass logistic models (MLM with ``baseline'' link), cumulative logit models (MLM with ``cumulative'' link), adjacent-categories logit models (MLM with ``adjacent'' link), and continuation-ratio logit models (MLM with ``continuation'' link). Its key functions and structure are displayed in Figure~\ref{fig:structure}. 

More specifically, to construct locally D-optimal designs, {\bf ForLion} provides two key functions by implementing the ForLion algorithm \citep{huang2024forlion}: 
\begin{itemize} 
\item \texttt{ForLion\_MLM\_Optimal} for finding locally D-optimal approximate designs under an MLM. 
\item \texttt{ForLion\_GLM\_Optimal} for finding locally D-optimal approximate designs under a GLM, which covers LM with ``identity'' link as a special case.
\end{itemize} 
The specialized arguments and their descriptions for these functions are summarized in  Table~\ref{tab:Special_arguments_for_the_ForLion_functions} of the Supplementary Material (Section~\ref{sec:function_arguments}).

To construct robust D-optimal designs against unknown parameter values, {\bf ForLion} package offers two key functions as well by implementing the EW ForLion algorithm \citep{lin2025ew}: 
\begin{itemize} 
\item \texttt{EW\_ForLion\_MLM\_Optimal} for finding EW D-optimal approximate designs under an MLM. 
\item \texttt{EW\_ForLion\_GLM\_Optimal} for finding EW D-optimal approximate designs under a GLM. \end{itemize} 
Details about additional arguments for implementing these functions are listed in  Table~\ref{tab:Special_arguments_for_the_EW_ForLion_functions} of the Supplementary Material (Section~\ref{sec:function_arguments}).

To obtain exact designs with user-defined grid points and number of experimental units from  approximate designs, {\bf ForLion} package provides the following two functions by implementing the rounding algorithm proposed by \cite{lin2025ew}:
\begin{itemize} 
\item \texttt{MLM\_Exact\_Design} for obtaining exact designs under an MLM. 
\item \texttt{GLM\_Exact\_Design} for obtaining exact designs under a GLM. 
\end{itemize}
A list of arguments commonly used in all the key functions is presented in Table~\ref{tab:Comment_arguments} of the Supplementary Material (Section~\ref{sec:function_arguments}).

When finding robust designs under an MLM, such as a cumulative logit model, the feasible parameter space may not be rectangular \citep{bu2020, lin2025ew}, and the computation for integral-based EW D-optimality (see \eqref{eq:f_EW}) is much more difficult,  especially with a moderate or large number $J$ of response categories. In the {\bf ForLion} package, we adopt the sample-based EW D-optimality (see \eqref{eq:f_SEW}) for MLMs. As for GLMs, we allow users to choose integral-based or sample-based D-optimality for deriving robust approximate and exact designs (see also Section~\ref{sec:EW_ForLion}). 

\section{Examples} \label{sec:example}

In this section, we demonstrate by examples how to use the {\bf ForLion} package to obtain locally D-optimal approximate designs, EW D-optimal approximate designs according to integral-based or sample-based EW D-optimality, and the corresponding exact designs, under different scenarios. The package vignette also includes a fully reproducible MLM example with five continuous factors based on a minimizing surface defects experiment. Further discussions for this experiment can be found in Section~S.3 in the Supplementary Material of \cite{huang2024forlion} for D-optimal designs, and Section~S5.1 in the Supplementary Material of \cite{lin2025ew} for EW D-optimal designs.

\subsection{An MLM example: Emergence of house flies experiment}\label{sec:fly_example}

We demonstrate the implementation of the \texttt{ForLion\_MLM\_Optimal()} function by exploring an emergence of house flies experiment. The original experiment, described by \cite{itepan1995}, involved $N = 3,500$ pupae uniformly assigned to $m=7$ different levels of a gamma radiation device (in Gy): $x_i = 80, \ 100,\ 120, \ 140, \ 160, \ 180,\  200$. The study recorded three categorical outcomes, namely \texttt{unopened}, \texttt{opened but died}, and \texttt{opened and emerged}, which clearly have an order or structure. A continuation-ratio non-proportional odds (npo) model with $J=3$, as a special case of MLM, was considered in the literature \citep{atkinson1999, bu2020, ai2023locally}:
\begin{eqnarray*}
    \log\left(\frac{\pi_{i1}}{\pi_{i2} + \pi_{i3}}\right) &=& \beta_{11} + \beta_{12} x_i + \beta_{13} x_i^2\ ,\\ 
\log\left(\frac{\pi_{i2}}{\pi_{i3}}\right) &=& \beta_{21} + \beta_{22} x_i\ ,
\end{eqnarray*}
with parameters $\hat{\boldsymbol\theta} = (\hat\beta_{11}, \hat\beta_{12}, \hat\beta_{13}, \hat\beta_{21}, \hat\beta_{22})^\top = (-1.935, -0.02642, 0.0003174,-9.159,$ $0.06386)^\top$ fitted from the pilot study by \cite{itepan1995}, where $i=1, \ldots, m$.

Following \cite{ai2023locally} and \cite{huang2024forlion}, we regard $\hat{\boldsymbol{\theta}}$ as the true parameter values and reconsider the experiment with a continuous range for the gamma radiation levels, namely $x_i \in \mathcal{X} = [0, 200]$. In this case, the $J \times p$ model matrix $\mathbf{X}_x$ (here $p=5$) and its derivative with respect to the continuous variable $x$ are 
\[
\mathbf{X}_x
=
\left(
\begin{array}{ccccc}
1 & x & x^2 & 0 & 0 \\
0 & 0 & 0   & 1 & x \\
0 & 0 & 0   & 0 & 0
\end{array}
\right)_{3\times 5}\ ,\>\>\>
\frac{\partial \mathbf{X}_x}{\partial x}
=
\left(
\begin{array}{ccccc}
0 & 1 & 2x & 0 & 0 \\
0 & 0 & 0  & 0 & 1 \\
0 & 0 & 0  & 0 & 0
\end{array}
\right)_{3\times 5}\ ,
\]
respectively. Then $\boldsymbol{\eta}(x) = \mathbf{X}_x \boldsymbol{\theta}=\left(\beta_{11}+\beta_{12} x+\beta_{13} x^2,\ \beta_{21}+\beta_{22} x,\ 0\right)^T$. Here the third row of the model matrix $\mathbf{X}_x$ consists entirely of structural zeros.  We keep this row to maintain a fixed $J \times p$ output dimension in our general implementation.

\medskip
\noindent
{\bf Finding a locally D-optimal approximate design}
\smallskip

\noindent
We start by defining the assumed model parameter values \texttt{theta}, the design matrix function  \texttt{hfunc.temp} for $\mathbf{X}_x$~, and the corresponding derivative function \texttt{hprime.temp} for $\partial \mathbf{X}_x/\partial x$. The function \texttt{hprime.temp} returns a list format to accommodate multiple continuous factors. In this example, the list contains a single matrix.
\footnotesize
\begin{verbatim}
> theta <- c(-1.935, -0.02642, 0.0003174, -9.159, 0.06386)
> hfunc.temp <- function(x){
+               matrix(data = c(1, x, x*x, 0, 0,
+                               0, 0, 0, 1, x,
+                               0, 0, 0, 0, 0), nrow = 3, ncol = 5, byrow = TRUE)} 
> hprime.temp <- function(x){
+                list(matrix_1 = matrix(data = c(0, 1, 2*x, 0, 0,
+                                                0, 0, 0, 0, 1,
+                                                0, 0, 0, 0, 0), 
+                                                nrow = 3, ncol = 5, byrow = TRUE))}
\end{verbatim}
\normalsize

Next, the \verb|ForLion_MLM_Optimal()| function can be applied to find a locally D-optimal approximate design under the continuation-ratio npo model using the following R code:
\footnotesize
\begin{verbatim}
> set.seed(123)
> forlion_MLM <- ForLion_MLM_Optimal(J = 3, n.factor = c(0),
+                factor.level = list(c(0, 200)), hfunc = hfunc.temp, 
+                h.prime = hprime.temp, bvec = theta, link = "continuation", 
+                Fi.func = Fi_MLM_func, delta0 = 1e-6, epsilon = 1e-12, 
+                reltol = 1e-8, delta = 0.15, maxit = 1000, random = TRUE, 
+                nram = 3, random.initial = TRUE, nram.initial = 3)
\end{verbatim}
\normalsize

%\sloppy
For the function arguments, we set \verb|n.factor = c(0)|, where 0 denotes a continuous factor, indicating that the experiment involves a single continuous factor and no discrete factors, and the range of the continuous factor is defined through \texttt{factor.level = list(c(0, 200))}, which is a closed interval. The obtained approximate design by the ForLion algorithm can be summarized and output using the \verb|print()| function as below:
\begin{verbatim}
> print(forlion_MLM)
Design Output
=========================== 
Count  X1        Allocation
--------------------------- 
1      103.5300  0.3981
2        0.0000  0.2027
3      149.2116  0.3992
=========================== 
m:
[1] 3
det:
[1] 54016299
convergence:
[1] TRUE
min.diff:
[1] 45.6816
x.close:
[1] 103.5300 149.2116
itmax:
[1] 23
\end{verbatim}

The above \verb|Design Output| identifies three design points $x_1 = 0.0000$, $x_2 = 103.5300$, and $x_3 = 149.2116$, along with the approximate allocations  $w_1 = 0.2027$, $w_2 = 0.3981$, and $w_3 = 0.3992$. Here \verb|m=3| matches the number of distinct design points. The determinant {\tt det} of the Fisher information matrix associated with the obtained design $\boldsymbol{\xi} = \{(x_i, w_i) \mid i=1,2,3\}$ is $\left| \mathbf F(\boldsymbol\xi, \hat{\boldsymbol{\theta}})\right|=54,016,299$. Furthermore, \verb|convergence: TRUE| indicates that the algorithm converged successfully. The minimum Euclidean distance {\tt min.diff} among the distinct design points is $45.6816$, and the closest pair {\tt x.close} of the design points is $103.5300$ and $149.2116$. Finally, \verb|itmax = 23| indicates that the number of iterations spent by the algorithm is $23$. 

To assess the variability of the obtained locally D-optimal design against the random seed prespecified, we rerun the algorithm with 10 randomly generated random seeds, while keeping all other settings fixed. In terms of relative efficiencies, the obtained designs are fairly stable (see Section~\ref{sec:assessing_sensitivity} in the Supplementary Material).

\medskip
\noindent
{\bf Obtaining exact designs from the locally D-optimal approximate design $\boldsymbol{\xi}$}
\smallskip

\noindent
Among the three reported design points in $\boldsymbol{\xi}$, two of them, namely $x_2=103.5300$ and $x_3 = 149.2116$ in Gy, may not be feasible in practice due to the sensitivity level of the gamma radiation device. For example, if the device can only allow the gamma ratiation level to be set as a multiple of 0.1 Gy, we may use the {\tt MLM\_Exact\_Design()} function with grid level $L=0.1$ as follows:
\footnotesize
\begin{Verbatim}
> forlion_MLM_exact <- MLM_Exact_Design(J = 3, k.continuous = 1, 
+                      design_x = forlion_MLM$x.factor, design_p = forlion_MLM$p,
+                      det.design = forlion_MLM$det, p = 5, ForLion = TRUE, 
+                      bvec = theta, delta2 = 1, L = 0.1, N = 3500,
+                      hfunc = hfunc.temp, link = "continuation")
\end{Verbatim}
\normalsize
where {\tt J = 3} represents the number of response categories, \verb|k.continuous = 1| indicates that there is only one continuous variable, namely   the gamma radiation level, \verb|design_x| is the set of design points in $\boldsymbol{\xi}$, \verb|design_p| stores the corresponding approximate allocations, \verb|det.design| is the maximized determinant of the Fisher information matrix, {\tt p = 5} stands for the number of parameters, and {\tt N = 3500} is the total number of experimental units. 
Furthermore, The argument {\tt ForLion = TRUE} indicates that the approximate design is obtained by the ForLion algorithm, while  \verb|ForLion = FALSE| corresponds to the EW ForLion algorithm. 
The converted exact design, denoted by $\boldsymbol{\xi}_{\rm exact}$ is output as below: 
\begin{verbatim}
> print(forlion_MLM_exact)
Design Output
=========================== 
Count  X1        Allocation
--------------------------- 
1      103.5000  0.3981
2        0.0000  0.2027
3      149.2000  0.3992
=========================== 
ni.design:
[1] 1393  710 1397
det:
[1] 54016013
rel.efficiency:
[1] 0.9999989
\end{verbatim}

The output of the exact design $\boldsymbol{\xi}_{\rm exact}$ shows that it contains three design points, namely $x_1=0$, $x_2=103.5$, $x_3=149.2$, along with the corresponding integer-valued allocations {\tt ni.design}, namely $n_1=710$, $n_2=1393$,  $n_3=1397$. The relative efficiency \verb|rel.efficiency| of $\boldsymbol{\xi}_{\rm exact}$ with respect to the D-optimal approximate design $\boldsymbol{\xi}$, is $(|{\mathbf F}(\boldsymbol{\xi}_{\rm exact}, \hat{\boldsymbol\theta})|/|{\mathbf F}(\boldsymbol{\xi}, \hat{\boldsymbol\theta})|)^{1/p} = 0.9999989$, or $99.99989\%$, which is highly efficient.

For illustration purposes, we repeat the procedure for some other grid levels, namely $L=1, \ 5, \ 10$, and $20$ as well. The corresponding exact designs, as well as $L=0.1$, are summarized in Table~\ref{tab:the_exact_designs_for_the_house_files}.
According to Table~\ref{tab:the_exact_designs_for_the_house_files}, the relative efficiency decreases as the rounding level $L$ increases. In practice, the experimenters may choose an appropriate $L$ as a compromise between the experimental requirements and the relative efficiency.

\begin{table}[htb]
    \centering
    \caption{Exact designs for house files experiment with different rounding levels and $N=3,500$}
    \resizebox{0.8\textwidth}{!}{%
    \begin{tabular}{crrrrrrrrrr}
    \toprule
    \multicolumn{1}{c}{} 
      & \multicolumn{2}{c}{\textbf{L = 0.1}}
      & \multicolumn{2}{c}{\textbf{L = 1}}
      & \multicolumn{2}{c}{\textbf{L = 5}}
      & \multicolumn{2}{c}{\textbf{L = 10}}
      & \multicolumn{2}{c}{\textbf{L = 20}} \\
    \cmidrule(lr){2-3} \cmidrule(lr){4-5} \cmidrule(lr){6-7} \cmidrule(lr){8-9} \cmidrule(lr){10-11}
   $i$ & $x_i$ & $n_i$ 
    & $x_i$ & $n_i$
    & $x_i$ & $n_i$ 
    & $x_i$ & $n_i$ 
    & $x_i$ & $n_i$ \\
    \midrule
    % -- Your data rows go here --
 1& 0 & 710 &  0 &  710 &  0&  710 &  0 & 710  & 0 & 710\\
 2 &  103.5& 1393 & 104  &  1393 & 105 & 1393  &  100 &  1393 & 100 & 1393\\
 3 & 149.2 & 1397 & 149  &  1397 & 150  &  1397 & 150  &  1397 &  140& 1397\\
  \midrule
Rel. Efficiency & \multicolumn{2}{c}{99.99989\%} & \multicolumn{2}{c}{99.98448\%} & \multicolumn{2}{c}{99.93424\%} & \multicolumn{2}{c}{99.48902\%} & \multicolumn{2}{c}{94.65724\%} \\
    \bottomrule
    \end{tabular}}
    \label{tab:the_exact_designs_for_the_house_files}
\end{table}

\medskip
\noindent
{\bf Finding a sample-based EW D-optimal approximate design}
\smallskip

\noindent
Next, we look for an EW D-optimal approximate design rather than a locally D-optimal one with the prespecified  $\hat{\boldsymbol{\theta}}$. It is more robust against misspecified parameter values \citep{lin2025ew}. More specifically, we first draw $B=1,000$ bootstrapped samples from the previous data (Table~1 in \cite{atkinson1999}), and then obtain the corresponding parameter estimates $\hat{\boldsymbol{\theta}}_j$ from the $j$th bootstrapped sample, for $j = 1, \ldots, B$. 
The $1,000$ parameter vectors are stored as matrix \verb|theta_matrix| as below:
\footnotesize
\begin{verbatim}
## simulate multinomial counts using the observed proportions as probabilities
> n <- 1000   # number of simulated datasets
> Ni <- 500   # multinomial sample size at each design point
> set.seed(2024)
## 7 design points with covariates (x1,x2), where x2 = x1^2
> x1_vec <- seq(80, 200, by = 20)
> x2_vec <- x1_vec^2
## Multinomial probabilities at each design point (rows sum to 1)
> prob_mat <- rbind(c( 62,  5, 433),
+                   c( 94, 24, 382),
+                   c(179, 60, 261),
+                   c(335, 80,  85),
+                   c(432, 46,  22),
+                   c(487, 11,   2),
+                   c(498,  2,   0)) / Ni
## Step 1: generate n simulated datasets with the specified probabilities; 
## each dataset has 7 rows (one per design point)
## sim_data[ , , k] is the k-th simulated dataset (7 x 5 matrix)
## columns: x1, x2, y1, y2, y3
> sim_data <- array(NA, dim = c(7, 5, n),
+                   dimnames = list(NULL, c("x1", "x2", "y1", "y2", "y3"), NULL))
> for (i in 1:7) {
+     Y_mat <- t(rmultinom(n, size = Ni, prob = prob_mat[i, ]))    # n x 3
+     Allsimdata_i <- cbind(x1 = x1_vec[i], x2 = x2_vec[i], Y_mat) # n x 5
+     for(k in 1:n){
+        sim_data[i, ,k] <- Allsimdata_i[k, ]
+    }
+  }
## Step 2: fit models for each simulated dataset and store selected coefficients
> theta_matrix <- matrix(0, nrow = n, ncol = 5)
> for (k in 1:n) {
+     data_k <- as.data.frame(sim_data[ , , k])
+     ## continuation-ratio model (VGAM: vglm, family = sratio)
+     ## fit1: predictors x1 + x2; fit2: predictor x1 only
+     fit1 <- vglm(cbind(y1, y2, y3) ~ x1 + x2, family = sratio, data = data_k)
+     fit2 <- vglm(cbind(y1, y2, y3) ~ x1,      family = sratio, data = data_k)
+     theta1 <- coef(fit1)
+     theta2 <- coef(fit2)
+     ## store selected coefficients 
+     ## The indices (1,3,5) and (2,4) follow coefficient ordering for family=sratio 
+     theta_matrix[k, ] <- c(theta1[c(1, 3, 5)], theta2[c(2, 4)])
}
\end{verbatim}
\normalsize

With the parameter matrix \verb|theta_matrix|, the functions \texttt{hfunc.temp} and \texttt{hprime.temp} previously defined, we use \verb|EW_ForLion_MLM_Optimal()| function to find a sample-based EW D-optimal approximate design as follows:
\footnotesize
\begin{verbatim}
> set.seed(123)
> ew_forlion_MLM <- EW_ForLion_MLM_Optimal(J = 3 ,n.factor = c(0),
+                   factor.level = list(c(0, 200)), hfunc = hfunc.temp, 
+                   h.prime = hprime.temp, bvec_matrix = theta_matrix, 
+                   link = "continuation", EW_Fi.func = EW_Fi_MLM_func, 
+                   delta0 = 1e-6, epsilon = 1e-12, reltol = 1e-8, delta = 0.15, 
+                   maxit = 1000, random = TRUE, nram = 1, random.initial = TRUE, 
+                   nram.initial = 3)
\end{verbatim}
\normalsize
The sample-based EW D-optimal approximate design, denoted by $\boldsymbol{\xi}_{\rm SEW}$, is output as below: 
\begin{verbatim}
> print(ew_forlion_MLM)
Design Output
=========================== 
Count  X1        Allocation
--------------------------- 
1        0.0000  0.2029
2      103.5039  0.3543
3      103.2826  0.0436
4      149.1144  0.3991
=========================== 
m:
[1] 4
det:
[1] 58719194
convergence:
[1] TRUE
min.diff:
[1] 0.2213
x.close:
[1] 103.5039 103.2826
itmax:
[1] 20
\end{verbatim}

For illustration purpose, we adopt $\delta=0.15$ as the merging threshold. The above \verb|Design Output| shows that the reported EW D-optimal approximate design contains four design points, namely $0.0000$, $103.2826$, $103.5039$, $149.1144$, and their corresponding approximate allocations are $0.2029$, $0.0436$, $0.3543$, $0.3991$, respectively. The determinant of the expected Fisher information matrix is $| \hat{E}\{{\mathbf F}({\boldsymbol \xi}_{\rm SEW}, \boldsymbol{\Theta})\}|=58,719,194$. 

\medskip
\noindent
{\bf Obtaining an exact design from the sample-based EW D-optimal design ${\boldsymbol{\xi}}_{\rm SEW}$}
\smallskip

\noindent
Similarly to the locally D-optimal approximate design $\boldsymbol{\xi}$, in practice we need to convert the sample-based EW D-optimal approximate design ${\boldsymbol{\xi}}_{\rm SEW}$ into an exact design with prespecified grid level $L$ and the total number $N$ of experimental units. We may use the same function \verb|MLM_Exact_Design()|. Instead of entering \verb|ForLion = TRUE| and the parameter vector \verb|bvec|, we need to set \verb|ForLion = FALSE| indicating EW ForLion algorithm, and input the bootstrapped parameter matrix \verb|theta_matrix| for argument \verb|bvec_matrix|. The corresponding exact design with $L = 0.1$ and $N = 3,500$ is obtained by the following R code:
\footnotesize
\begin{Verbatim}
> ew_forlion_MLM_exact <- MLM_Exact_Design(J = 3, k.continuous = 1, 
+                          design_x = ew_forlion_MLM$x.factor,
+                          design_p = ew_forlion_MLM$p, 
+                          det.design = ew_forlion_MLM$det, p = 5, ForLion = FALSE,
+                          bvec_matrix = theta_matrix, delta2 = 1, L = 0.1, 
+                          N = 3500, hfunc = hfunc.temp, link = "continuation")
\end{Verbatim}
\normalsize
The obtained exact design,  denoted by ${\boldsymbol{\xi}}'_{\rm exact}$, is output as below:
\begin{verbatim}
> print(ew_forlion_MLM_exact)
Design Output
=========================== 
Count  X1        Allocation
--------------------------- 
1        0.0000  0.2029
2      149.1000  0.3991
3      103.5000  0.3980
=========================== 
ni.design:
[1]  710 1397 1393
det:
[1] 58718854
rel.efficiency:
[1] 0.9999988
\end{verbatim}

Instead of four design points in ${\boldsymbol{\xi}}_{\rm SEW}$, the exact design ${\boldsymbol{\xi}}'_{\rm exact}$ contains only three design points, namely $0, 103.5$, and $149.1$, along with the integer-valued allocations  $710, 1393$, and $1397$. The relative efficiency of ${\boldsymbol{\xi}}'_{\rm exact}$ with respect to ${\boldsymbol{\xi}}_{\rm SEW}$ is \[
(|\hat{E}\{{\mathbf F}({\boldsymbol{\xi}}'_{\rm exact}, \boldsymbol \Theta)\}|/|\hat{E}\{{\mathbf F}({\boldsymbol{\xi}}_{\rm SEW}, \boldsymbol\Theta)\}|)^{1/p} = 0.9999988\mbox{ or }99.99988\%\ .
\]

\subsection{A GLM example: Electrostatic discharge (ESD) experiment}\label{sec:ESD_example}

\cite{lukemire2018} revisited the electrostatic discharge (ESD) experiment, initially described by \cite{whitman2006}, as a GLM example (see also \cite{huang2024forlion} and \cite{lin2025ew}). It involves a binary response, whether a certain part of the semiconductor fails, and five mixed experimental factors. The first four factors, namely \texttt{LotA} ($x_1$), \texttt{LotB} ($x_2$), \texttt{ESD} ($x_3$), and \texttt{Pulse} ($x_4$), take values in $\{-1,1\}$, while the fifth factor \texttt{Voltage} ($x_5$) is continuous within the range $[25, 45]$. Let $\mathbf{x}=(x_1,\ldots,x_5)^\top$ denote the factor vector.
A logistic model is considered for this experiment: 
\[
 \text{logit}(\mu)=\beta_0+\beta_1 x_1+\beta_2 x_2+\beta_3 x_3+\beta_4 x_4+\beta_5x_5+\beta_{34}(x_3 \times x_4) \ .
 \]

For implementation in our algorithm, we put the coefficient of the continuous factor $\beta_5$ first, and the intercept last. Accordingly, we reorder the parameter vector as $\boldsymbol \theta = (\beta_5, \beta_1, \beta_2, \beta_3, \beta_4, \beta_{34}, \beta_0)^\top$ and define the corresponding predictor vector
$
\mathbf{h}(\mathbf{x})=\bigl(x_5,\ x_1,\ x_2,\ x_3,\ x_4,\ x_3 \times x_4,\ 1\bigr)^\top
$.
Letting $\mathbf{x}_{(1)}=x_5$ denote the continuous variable, the corresponding derivative used by our algorithm is 
$
\partial \mathbf{h}(\mathbf{x})/\partial \mathbf{x}^\top_{(1)}=\bigl(1,0,0,0,0,0,0\bigr)^\top
$.
Here $\partial \mathbf{h}(\mathbf{x})/\partial \mathbf{x}_{(1)}^{\top}$ is in general a $p \times k$ matrix, where $p$ is the number of parameters, and $k$ is the number of continuous factors (for illustration purposes, in Section~\ref{sec:additional_example_ESD} of the Supplementary Material, we consider a different model involving {\tt Pulse} with three levels $\{-1,0,1\}$ and an interaction between $x_4$ and the continuous factor $x_5$).

\medskip
\noindent
{\bf Finding a locally D-optimal approximate design}
\smallskip

\noindent
To use \verb|ForLion_GLM_Optimal()| function, we first need to specify the function for generating the design matrix, and the assumed model parameters $\boldsymbol{\theta} = (0.35, 1.50, -0.2, -0.15,$ $0.25, 0.4, -7.5)^\top$ adopted by \cite{lukemire2018} and \cite{huang2024forlion}:
\begin{verbatim}
## After reordering the components in x: x = (x5, x1, x2, x3, x4)^T
## x -> h(x) = (x5, x1, x2, x3, x4, x3*x4, 1)^T
>  hfunc.temp <- function(x) {c(x, x[4]*x[5], 1);};  
>  beta.value <- c(0.35, 1.50, -0.2, -0.15, 0.25, 0.4, -7.5)
>  variable_names <- c("Vol.", "LotA", "LotB", "ESD", "Pul.")
## Using self defined function for the dh(x)/d(x)
> hprime.temp <- function(x){
+                matrix_1 = matrix(data = c(1, 0, 0, 0, 0, 0, 0),
+                                  nrow = 7, ncol = 1, byrow = TRUE)
}
\end{verbatim}

Next, we use the argument \texttt{n.factor = c(0, 2, 2, 2, 2)} to specify the experimental factor structure, where {\tt 0} represents the continuous factor {\tt Voltage} (always come first) and the subsequent $2$'s represent binary discrete factors (four in total). The corresponding list of factor levels are provided in \texttt{factor.level = list(c(25,45), c(-1,1), c(-1,1),$ $c(-1,1), c(-1,1))}, where {\tt c(25,45)} stands for an interval, $[25, 45]$, for the continuous factor, and {\tt c(-1,1)} represents a set, $\{-1, 1\}$, for a discrete factor. The corresponding R commands are listed as below:
\footnotesize
\begin{verbatim}
>  set.seed(482)
>  forlion_GLM <- ForLion_GLM_Optimal(n.factor = c(0, 2, 2, 2, 2), 
+                 factor.level = list(c(25, 45), c(-1, 1), c(-1, 1), c(-1, 1), 
+                 c(-1, 1)), var_names = variable_names, hfunc = hfunc.temp, 
+                 h.prime = hprime.temp, bvec = beta.value, link = "logit", 
+                 delta0 = 1e-5, epsilon = 1e-12, reltol = 1e-7, random = TRUE, 
+                 nram = 1, random.initial = TRUE, nram.initial = 1, delta = 0.01, 
+                 maxit = 1000, logscale = TRUE)
\end{verbatim}
\normalsize

The results obtained from the GLM-adapted ForLion algorithm can be summarized and displayed using the \texttt{print()} function. It provides a concise overview of the characteristics of the obtained optimal design, including the selected design points, their corresponding allocations, the determinant of the Fisher information matrix, etc. 
\begin{verbatim}
> print(forlion_GLM)
Design Output
============================================================== 
Count  Vol.     LotA     LotB     ESD      Pul.     Allocation
-------------------------------------------------------------- 
1      25.0000  -1.0000  -1.0000   1.0000  -1.0000  0.1165
2      27.5443  -1.0000  -1.0000  -1.0000  -1.0000  0.0156
3      25.0000  -1.0000   1.0000  -1.0000  -1.0000  0.0895
4      32.7748  -1.0000   1.0000   1.0000  -1.0000  0.1313
5      25.0000  -1.0000  -1.0000   1.0000   1.0000  0.0854
6      25.0000   1.0000   1.0000   1.0000  -1.0000  0.1331
7      25.0000  -1.0000   1.0000   1.0000   1.0000  0.0922
8      25.0000   1.0000  -1.0000   1.0000  -1.0000  0.0136
9      25.0000  -1.0000   1.0000   1.0000  -1.0000  0.0341
10     29.0549  -1.0000   1.0000  -1.0000  -1.0000  0.0042
11     25.0000  -1.0000  -1.0000  -1.0000   1.0000  0.0367
12     25.0000  -1.0000  -1.0000  -1.0000  -1.0000  0.0748
13     28.6912  -1.0000  -1.0000  -1.0000   1.0000  0.0722
14     25.0000  -1.0000   1.0000  -1.0000   1.0000  0.1008
============================================================== 
m:
[1] 14
det:
[1] 1.268957e-05
convergence:
[1] TRUE
min.diff:
[1] 2
x.close:
     [,1] [,2] [,3] [,4] [,5]
[1,]   25   -1   -1    1   -1
[2,]   25   -1   -1    1    1
itmax:
[1] 298
\end{verbatim}

The above \verb|Design Output| shows that the obtained design, denoted by $\boldsymbol{\xi}$, contains $14$ design points in the table,  whose levels are listed in the same order as the ones specified by {\tt factor.level}. The last column of the table is the corresponding approximate allocations for the design points. The determinant of the Fisher information matrix, namely {\tt det}, is $\left| \mathbf F(\boldsymbol \xi, \boldsymbol{\theta})\right| = 1.268957 \times 10^{-5}$.  In this locally D-optimal approximate design, the minimum Euclidean distance between the design points is equal to $2$, and the closest pair of design points is reported by \texttt{x.close}. The algorithm converges ({\tt convergence:TRUE}) with the number of iterations \texttt{itmax = 298}.

\medskip
\noindent
{\bf Obtaining exact designs based on the locally D-optimal approximate design $\boldsymbol{\xi}$}
\smallskip

\noindent
The approximate design $\boldsymbol{\xi}$ specifies accurate levels of the continuous factor \texttt{Voltage} like $27.5443$, $28.6912$, $29.0549$, and $32.7748$. Those voltage levels may not be able to be maintained precisely in the experiment.  If, for example, the voltage in this experiment can only be controlled to be a multiple of $L=0.1$, we may employ \verb|GLM_Exact_Design()| function to convert $\boldsymbol{\xi}$ into a feasible exact design with modified voltage levels and integer-valued allocations.
For illustration purpose, we set the total number of observations to $N = 500$, the merging threshold \verb|delta2| to $0.5$, and the rounding level to $L = 0.1$ for the only continuous factor \texttt{Voltage}. The corresponding exact design can be obtained by the following R code:  
\footnotesize
\begin{Verbatim}
> forlion_GLM_exact <- GLM_Exact_Design(k.continuous = 1, 
+                      design_x = forlion_GLM$x.factor, design_p = forlion_GLM$p, 
+                      var_names = variable_names, det.design = forlion_GLM$det, 
+                      p = 7, ForLion = TRUE, bvec = beta.value, delta2 = 0.5, 
+                      L = 0.1, N = 500, hfunc = hfunc.temp, link = "logit")
\end{Verbatim}
\begin{verbatim}
> print(forlion_GLM_exact)
Design Output
============================================================== 
Count  Vol.     LotA     LotB     ESD      Pul.     Allocation
-------------------------------------------------------------- 
1      25.0000  -1.0000  -1.0000   1.0000  -1.0000  0.1165
2      27.5000  -1.0000  -1.0000  -1.0000  -1.0000  0.0156
3      25.0000  -1.0000   1.0000  -1.0000  -1.0000  0.0895
4      32.8000  -1.0000   1.0000   1.0000  -1.0000  0.1313
5      25.0000  -1.0000  -1.0000   1.0000   1.0000  0.0854
6      25.0000   1.0000   1.0000   1.0000  -1.0000  0.1331
7      25.0000  -1.0000   1.0000   1.0000   1.0000  0.0922
8      25.0000   1.0000  -1.0000   1.0000  -1.0000  0.0136
9      25.0000  -1.0000   1.0000   1.0000  -1.0000  0.0341
10     29.1000  -1.0000   1.0000  -1.0000  -1.0000  0.0042
11     25.0000  -1.0000  -1.0000  -1.0000   1.0000  0.0367
12     25.0000  -1.0000  -1.0000  -1.0000  -1.0000  0.0748
13     28.7000  -1.0000  -1.0000  -1.0000   1.0000  0.0722
14     25.0000  -1.0000   1.0000  -1.0000   1.0000  0.1008
============================================================== 
ni.design:
[1] 58  8 45 66 43 67 46  7 17  2 18 37 36 50
det:
[1] 1.268788e-05
rel.efficiency:
[1] 0.999981
\end{verbatim}
\normalsize

The \verb|Design Output| shows that the obtained exact design, denoted by $\boldsymbol{\xi}_{\rm exact}$, only contains $14$ design points as listed in the table of design output. The levels of {\tt Voltage} (listed as the first factor in the table) have been rounded to multiples of $L = 0.1$. With $N = 500$, the integer-values allocations are listed in {\tt ni.design}. 
The relative efficiency of the exact design $\boldsymbol{\xi}_{\rm exact}$ with respect to the approximate design $\boldsymbol{\xi}$ is provided as {\tt rel.efficiency}, that is, $(|{\mathbf F}(\boldsymbol{\xi}_{\rm exact}, \boldsymbol \theta)|/|{\mathbf F}(\boldsymbol{\xi}, \boldsymbol \theta)|)^{1/p} = 0.999981$ or $99.9981\%$.

To illustrate how the exact design changes along with $N$, we generate two more exact designs with \verb|N = 100|, \verb|N = 500|, respectively,  both with \verb|L = 0.5|. Both designs are listed in  Table~\ref{tab:The_final_exact_designs_for_the_ESD_experiment}. Their relative efficiencies are $99.92635\%$ for \verb|N = 100| and $99.98025\%$ for \verb|N = 500|.

\begin{table}[htbp]
    \centering
    \caption{Exact designs with $L=0.5$ and different $N$'s for the ESD experiment}
    \resizebox{0.8\textwidth}{!}{
    \begin{tabular}{c|rrrrrr|c|rrrrrr} 
\hline Support & \multicolumn{6}{c|}{$N=100$}   &Support  &   \multicolumn{6}{c}{$N=500$}\\
point & Vol. & LotA & LotB & ESD&Pul.  &$n_i$ &point & Vol. & LotA & LotB & ESD&Pul. &$n_i$\\
\hline
1 &  25.0 &  -1  &  -1  &  1 &  -1 & 12  &
1 &  25.0 &  -1  &  -1  &  1 &  -1 &  58 \\
 
2 &  27.5 &  -1  &  -1  & -1 &  -1 &  2  &
2 &  27.5 &  -1  &  -1  & -1 &  -1 & 8 \\

3 &  25.0 &  -1  &   1  & -1 &  -1 &  9  &
3 &  25.0 &  -1  &   1  & -1 &  -1 &  45 \\

4 &  33.0 &  -1  &   1  &  1 &  -1 & 13   &
4 &  33.0 &  -1  &   1  &  1 &  -1 & 66   \\

5 &  25.0 &  -1  &  -1  &  1 &   1 &  9  &
5 &  25.0 &  -1  &  -1  &  1 &   1 & 43  \\

6 &  25.0 &   1  &  1   &  1 &  -1 &  13  &
6 &  25.0 &   1  &  1   &  1 &  -1 &  67\\

7 &  25.0 &  -1  &  1   &  1 &   1 &  9  &
7 &  25.0 &  -1  &  1   &  1 &   1 &  46\\

8 &  25.0 &   1  & -1   &  1 &  -1 &  1  &
8 &  25.0 &   1  & -1   &  1 &  -1 &  7 \\

9 &  25.0 &  -1  &   1  &  1 &  -1 &  3  &
9 &  25.0 &  -1  &   1  &  1 &  -1 &  17 \\

10 &  -  & -   &  -  & -   & -  & - &
10 &  29.0 &  -1 &  1  &  -1 &  -1 &  2 \\

11 &  25.0 &  -1 & -1  &  -1 &   1 &  4 &
11 &  25.0 &  -1 & -1  &  -1 &   1 &  18  \\

12 &  25.0 &  -1 & -1  & -1  &  -1 & 8  &
12 &  25.0 &  -1 & -1  & -1  &  -1 & 37  \\

13 &  28.5 &  -1 & -1  & -1  &   1 & 7  &
13 &  28.5 &  -1 & -1  & -1  &   1 & 36  \\

14 &  25.0 &  -1 &  1  & -1  &   1 &  10 &
14 &  25.0 &  -1 &  1  & -1  &   1 &  50  \\
\hline
\end{tabular}\label{tab:The_final_exact_designs_for_the_ESD_experiment}
}
\end{table}

\medskip
\noindent
{\bf Finding integral-based EW D-optimal approximate design}
\smallskip

\noindent
To find a robust D-optimal approximate design against possibly misspecified $\boldsymbol{\theta}$, we adopt the prior distribution suggested by 
\cite{huang2024forlion}. That is, we assume a prior distribution for $\boldsymbol{\Theta}$ (i.e., a randomized version of $\boldsymbol{\theta}$), such that, {\it (i)} all components of $\boldsymbol{\Theta}$ are independent of each other; and {\it (ii)} each component of $\boldsymbol{\Theta}$ follows a uniform distribution listed below:
$$\beta_0 \sim U(-8,-7),\quad \beta_1 \sim U(1,2), \quad \beta_2 \sim U(-0.3,-0.1), \quad  \beta_3 \sim U(-0.3,0), $$ 
$$ \beta_4 \sim U(0.1,0.4) , \quad \beta_5 \sim U(0.25, 0.45),\quad \beta_{34} \sim U(0.35,0.45).$$ 

To find an integral-based EW D-optimal approximate design given the prior distribution above, we first use two vectors, {\tt paras\_lowerbound} and {\tt paras\_upperbound}, to denote the lower bounds and upper bounds of the uniform distributions, respectively. Note that the orders of vector coordinates must match the same order as in {\tt factor.level}. Then we can define the probability density function (pdf) of the prior distribution by \verb|gjoint| function  below: 

\begin{verbatim}
> paras_lowerbound <- c(0.25, 1, -0.3, -0.3, 0.1, 0.35, -8.0)
> paras_upperbound <- c(0.45, 2, -0.1,  0.0, 0.4, 0.45, -7.0)
## the prior distributions are uniform distributions
> gjoint_b <- function(x) {
+             Func_b = 1/(prod(paras_upperbound-paras_lowerbound))
+             return(Func_b)
}  
\end{verbatim}

By utilizing the \verb|EW_ForLion_GLM_Optimal()| function with specified arguments, we obtain an integral-based EW D-optimal design as below:
\footnotesize
\begin{verbatim}
> set.seed(482)
> ew_forlion_GLM <- EW_ForLion_GLM_Optimal(n.factor = c(0, 2, 2, 2, 2), 
+                   factor.level = list(c(25,45),c(-1,1),c(-1,1),c(-1,1),c(-1,1)), 
+                   var_names = variable_names, hfunc = hfunc.temp, 
+                   h.prime = hprime.temp, Integral_based = TRUE, 
+                   joint_Func_b = gjoint_b, Lowerbounds = paras_lowerbound, 
+                   Upperbounds = paras_upperbound, link = "logit", delta0 = 1e-5, 
+                   epsilon = 1e-12, reltol = 1e-5, delta = 0.01, maxit = 500, 
+                   random = TRUE, nram = 1, logscale = TRUE)
\end{verbatim}

\begin{verbatim}
> print(ew_forlion_GLM)
Design Output
============================================================== 
Count  Vol.     LotA     LotB     ESD      Pul.     Allocation
-------------------------------------------------------------- 
1      25.0000  -1.0000  -1.0000  -1.0000   1.0000  0.0875
2      25.0000  -1.0000   1.0000   1.0000   1.0000  0.0845
3      25.0000  -1.0000  -1.0000  -1.0000  -1.0000  0.0848
4      25.0000   1.0000   1.0000  -1.0000   1.0000  0.0621
5      38.9047  -1.0000   1.0000   1.0000  -1.0000  0.0214
6      25.0000   1.0000   1.0000  -1.0000  -1.0000  0.0356
7      25.0000  -1.0000  -1.0000   1.0000   1.0000  0.0856
8      25.0000  -1.0000   1.0000  -1.0000   1.0000  0.0515
9      25.0000  -1.0000   1.0000  -1.0000  -1.0000  0.0690
10     33.1161  -1.0000   1.0000   1.0000   1.0000  0.0022
11     35.4140  -1.0000   1.0000  -1.0000   1.0000  0.0028
12     25.0000   1.0000   1.0000   1.0000  -1.0000  0.0443
13     25.0000   1.0000   1.0000   1.0000   1.0000  0.0090
14     35.3993  -1.0000   1.0000  -1.0000   1.0000  0.0352
15     25.0000  -1.0000   1.0000   1.0000  -1.0000  0.0901
16     25.0000   1.0000  -1.0000   1.0000  -1.0000  0.0743
17     34.0238  -1.0000   1.0000  -1.0000  -1.0000  0.0157
18     37.1975  -1.0000  -1.0000   1.0000  -1.0000  0.0455
19     25.0000  -1.0000  -1.0000   1.0000  -1.0000  0.0410
20     38.9522  -1.0000   1.0000   1.0000  -1.0000  0.0580
============================================================== 
m:
[1] 20
det:
[1] 4.552703e-06
convergence:
[1] TRUE
min.diff:
[1] 0.0147
x.close:
        [,1] [,2] [,3] [,4] [,5]
[1,] 35.4140   -1    1   -1    1
[2,] 35.3993   -1    1   -1    1
itmax:
[1] 56
\end{verbatim}
\normalsize

The reported integral-based EW D-optimal design, denoted by $\boldsymbol{\xi}_{\rm EW}$, consists of $20$ design points. The corresponding determinant {\tt det} of the expected Fisher information matrix is $| E\{{\mathbf F}({\boldsymbol \xi}_{\rm EW}, \boldsymbol{\Theta})\}| = 4.552703 \times 10^{-6}$. Having performed on a Windows 11 laptop with 32GB of RAM and a 13th Gen Intel Core i7-13700HX processor, with R version 4.4.2, the above procedure costs 2,865 seconds.

\medskip
\noindent
{\bf Obtaining an exact design based on the integral-based EW D-optimal design $\boldsymbol{\xi}_{\rm EW}$}
\smallskip

\noindent
Similarly to the locally D-optimal design $\boldsymbol{\xi}$, we also need to convert the EW D-optimal approximate design $\boldsymbol{\xi}_{\rm EW}$ into an exact design for practical uses. The same function \verb|GLM_Exact_Design()| can be applied, but with \verb|ForLion = FALSE| indicating that the original design was obtained by an EW ForLion algorithm. In this case, we also need to input the pdf \verb|joint_Func_b| for the prior distribution, along with the lower and upper bounds of ranges, namely \verb|Lowerbounds| and \verb|Upperbounds|. For illustration purpose, we still use the rounding level $L=0.1$ for the continuous factor and the total number of observations $N=500$.

\footnotesize
\begin{Verbatim}
> ew_forlion_exact <- GLM_Exact_Design(k.continuous = 1,
+                     design_x = ew_forlion_GLM$x.factor, 
+                     design_p = ew_forlion_GLM$p, var_names = variable_names, 
+                     det.design = ew_forlion_GLM$det, p = 7, ForLion = FALSE, 
+                     Integral_based = TRUE, joint_Func_b = gjoint_b,
+                     Lowerbounds = paras_lowerbound, 
+                     Upperbounds = paras_upperbound, delta2 = 0.5, L = 0.1,
+                     N = 500, hfunc = hfunc.temp, link = "logit")
\end{Verbatim}
\begin{verbatim}
> print(ew_forlion_exact)
Design Output
============================================================== 
Count  Vol.     LotA     LotB     ESD      Pul.     Allocation
-------------------------------------------------------------- 
1      25.0000  -1.0000  -1.0000  -1.0000   1.0000  0.0875
2      25.0000  -1.0000   1.0000   1.0000   1.0000  0.0845
3      25.0000  -1.0000  -1.0000  -1.0000  -1.0000  0.0848
4      25.0000   1.0000   1.0000  -1.0000   1.0000  0.0621
5      25.0000   1.0000   1.0000  -1.0000  -1.0000  0.0356
6      25.0000  -1.0000  -1.0000   1.0000   1.0000  0.0856
7      25.0000  -1.0000   1.0000  -1.0000   1.0000  0.0515
8      25.0000  -1.0000   1.0000  -1.0000  -1.0000  0.0690
9      33.1000  -1.0000   1.0000   1.0000   1.0000  0.0022
10     25.0000   1.0000   1.0000   1.0000  -1.0000  0.0443
11     25.0000   1.0000   1.0000   1.0000   1.0000  0.0090
12     25.0000  -1.0000   1.0000   1.0000  -1.0000  0.0901
13     25.0000   1.0000  -1.0000   1.0000  -1.0000  0.0743
14     34.0000  -1.0000   1.0000  -1.0000  -1.0000  0.0157
15     37.2000  -1.0000  -1.0000   1.0000  -1.0000  0.0455
16     25.0000  -1.0000  -1.0000   1.0000  -1.0000  0.0410
17     35.4000  -1.0000   1.0000  -1.0000   1.0000  0.0380
18     38.9000  -1.0000   1.0000   1.0000  -1.0000  0.0794
============================================================== 
ni.design:
 [1] 44 42 42 31 18 43 26 34  1 22  4 45 37  8 23 21 19 40
det:
[1] 4.551996e-06
rel.efficiency:
[1] 0.9999778
\end{verbatim}
\normalsize

The reported exact design consists of $18$ design points, along with their corresponding allocations provided by {\tt ni.design}. Its relative efficiency with respect to $\boldsymbol{\xi}_{\rm EW}$ is $0.9999778$ or $99.99778\%$. 

\medskip
\noindent
{\bf Finding a sample-based EW D-optimal approximate design}
\smallskip

\noindent
Given the same prior distribution for obtaining $\boldsymbol{\xi}_{\rm EW}$~, we can also simulate $B=1,000$ random parameter vectors $\{\boldsymbol{\theta}_1, \ldots, \boldsymbol{\theta}_B\}$ from the prior distribution, and then find a sample-based EW D-optimal approximate design based on the simulated parameter vectors.

\begin{verbatim}
> nrun <- 1000
> set.seed(0713)
> b_0 <- runif(nrun, -8, -7)
> b_1 <- runif(nrun, 1, 2)
> b_2 <- runif(nrun, -0.3, -0.1)
> b_3 <- runif(nrun, -0.3, 0)
> b_4 <- runif(nrun, 0.1, 0.4)
> b_5 <- runif(nrun, 0.25, 0.45)
> b_34 <- runif(nrun, 0.35, 0.45)
> beta.matrix <- cbind(b_5,b_1,b_2,b_3,b_4,b_34,b_0)
\end{verbatim}

Similarly to obtaining $\boldsymbol{\xi}_{\rm EW}$~, we can also apply the \texttt{EW\_ForLion\_GLM\_Optimal()} function, but with \verb|Integral_based = FALSE| indicating sample-based EW D-optimality, and input the sampled parameter matrix \verb|beta.matrix| obtained above for argument \verb|b_matrix|.

\footnotesize
\begin{verbatim}
> set.seed(482)
> sample_ew_forlion_GLM <- EW_ForLion_GLM_Optimal(n.factor = c(0, 2, 2, 2, 2), 
+                          factor.level = list(c(25, 45), c(-1, 1), c(-1, 1),
+                          c(-1, 1), c(-1, 1)), var_names = variable_names, 
+                          hfunc = hfunc.temp, h.prime = hprime.temp, 
+                          Integral_based = FALSE, b_matrix = beta.matrix, 
+                          link = "logit", delta0 = 1e-5, epsilon = 1e-12, 
+                          reltol = 1e-6, delta = 0.01, maxit = 500, 
+                          random = TRUE, nram = 1, logscale = TRUE)
\end{verbatim}
\begin{verbatim}
> print(sample_ew_forlion_GLM)
Design Output
============================================================== 
Count  Vol.     LotA     LotB     ESD      Pul.     Allocation
-------------------------------------------------------------- 
1      25.0000  -1.0000  -1.0000   1.0000   1.0000  0.0851
2      25.0000  -1.0000   1.0000  -1.0000   1.0000  0.0723
3      33.5304  -1.0000   1.0000  -1.0000  -1.0000  0.0095
4      25.0000  -1.0000  -1.0000   1.0000  -1.0000  0.0640
5      25.0000   1.0000   1.0000   1.0000  -1.0000  0.0499
6      25.0000   1.0000   1.0000  -1.0000  -1.0000  0.0310
7      25.0000  -1.0000   1.0000   1.0000   1.0000  0.0882
8      25.0000  -1.0000   1.0000  -1.0000  -1.0000  0.0743
9      38.4919  -1.0000   1.0000   1.0000  -1.0000  0.1171
10     33.2875  -1.0000  -1.0000  -1.0000   1.0000  0.0403
11     25.0000   1.0000  -1.0000   1.0000  -1.0000  0.0702
12     25.0000  -1.0000  -1.0000  -1.0000  -1.0000  0.0843
13     25.0000  -1.0000   1.0000   1.0000  -1.0000  0.0738
14     25.0000   1.0000   1.0000  -1.0000   1.0000  0.0612
15     25.0000  -1.0000  -1.0000  -1.0000   1.0000  0.0660
16     36.7975  -1.0000  -1.0000   1.0000  -1.0000  0.0084
17     25.0000   1.0000   1.0000   1.0000   1.0000  0.0037
18     33.5593  -1.0000   1.0000  -1.0000  -1.0000  0.0008
============================================================== 
m:
[1] 18
det:
[1] 4.229431e-06
convergence:
[1] TRUE
min.diff:
[1] 0.0289
x.close:
        [,1] [,2] [,3] [,4] [,5]
[1,] 33.5304   -1    1   -1   -1
[2,] 33.5593   -1    1   -1   -1
itmax:
[1] 96
\end{verbatim}
\normalsize

The obtained sample-based EW D-optimal approximate design, denoted by $\boldsymbol{\xi}_{\rm SEW}$~, contains {\tt m} $=18$ design points, with {\tt det} $=$ $| \hat{E}\{{\mathbf F}({\boldsymbol \xi}_{\rm SEW}, \boldsymbol{\Theta})\}| = 4.229431 \times 10^{-6}$.

\medskip
\noindent
{\bf Comparing sample-based and integral-based EW D-optimal designs}
\smallskip

\noindent
According to a simulation study done by \cite{lin2025ew} on a minimizing surface defects experiment, sample-based EW D-optimal designs are fairly robust in terms of relative efficiencies against different set of simulated parameter vectors. 

In this paper, we use this ESD experiment to compare the integral-based EW D-optimal design $\boldsymbol{\xi}_{\rm EW}$ with sample-based EW D-optimal designs based on six different sets of simulated parameter vectors. More specifically, for $b=1$ (representing $B=100$) and $b=2$ (representing $B=1,000$), we simulate three random sets of parameter vectors of size $B$, labeled by $j=1,2,3$, and find the corresponding sample-based EW D-optimal designs, denoted by $\boldsymbol{\xi}_{bj}$~.
Then we calculate the relative efficiencies of $\boldsymbol{\xi}_{bj}$ with respect to $\boldsymbol{\xi}_{\rm EW}$, in terms of the integral-based EW D-optimality. 
That is,
\[
\left(\frac{|E\{{\mathbf F}(\boldsymbol{\xi}_{bj}, \boldsymbol{\Theta})\}|}{|E\{{\mathbf F}(\boldsymbol{\xi}_{\rm EW}, \boldsymbol{\Theta})\}|}\right)^{1/p}\ ,
\]
for $b = 1, \ 2$, $j = 1, \ 2,\ 3$, and $p = 7$. The relative efficiencies and the corresponding numbers of distinct design points contained in $\boldsymbol{\xi}_{bj}$ are shown in the left and right matrices below, respectively:
\[
\bordermatrix{
  & $j = 1$ & $j = 2$ & $j = 3$  \cr
$b = 1$ & 0.9982641 & 0.9975130 & 0.9989559 \cr
$b = 2$ & 0.9997052 & 0.9997704 & 0.9994012 \cr
}\ ,
\qquad
\bordermatrix{
  & $j = 1$ & $j = 2$ & $j = 3$  \cr
$b = 1$ & 17  & 21  & 20 \cr
$b = 2$ & 20  & 19  & 18 \cr
}\ .
\]
The minimum relative efficiency is $0.997513$ or $99.7513\%$, which is fairly high. The numbers of distinct points range from 17 to 21, which vary from design to design, but are not so different from each other.

%% -- Summary/conclusions/discussion -------------------------------------------
\section{Summary and Discussion} \label{sec:summary}

In this paper, we introduce the {\bf ForLion} package, which facilitates the users to find D-optimal or EW D-optimal designs of experiments involving discrete factors only,  continuous factors only, or mixed factors. In {\bf ForLion}, \texttt{factor.level} is used only to declare the design space, which enumerates the levels of qualitative factors and specifies the ranges of continuous factors. For a qualitative factor with $K \ge 3$ levels, the corresponding regression model can be constructed either by $K-1$ indicator variables or by other contrast coding choices through the user-defined \texttt{hfunc}. Interaction terms involving continuous factors can also be specified in \texttt{hfunc}, with the required derivative information provided via \texttt{h.prime} when needed (see Section~\ref{sec:additional_example_ESD} in the Supplementary Material for such an example). In the GLM setting, \texttt{h.prime} is not required for a main-effects model or for a model with interactions restricted to discrete factors. In this case, the needed derivatives are calculated automatically. In the MLM setting, if \texttt{h.prime} is not provided, the package uses numerical derivatives by default.

The current version 0.4.0 of the {\bf ForLion} package supports both GLM and MLM for various experimental scenarios. It is worth noting that a regular linear regression model is a special case of GLM with an identity link function, and is covered by the {\bf ForLion} package as well. In this package, the functions \texttt{ForLion\_MLM\_Optimal()} and \texttt{ForLion\_GLM\_Optimal()} can be used for determining locally D-optimal approximate designs, if the experimenter is certain about the values of the parameters. When the true parameter values are unknown, while either a prior distribution $Q(\cdot)$ on $\boldsymbol{\Theta}$ or a dataset from a pilot study is available, we provide functions \texttt{EW\_ForLion\_MLM\_Optimal()} and \texttt{EW\_ForLion\_GLM\_Optimal()} to find EW D-optimal approximate designs. 
Having obtained D-optimal approximate designs, we also provide functions \texttt{GLM\_Exact\_Design()} for GLM and \texttt{MLM\_Exact\_Design()} for MLM to convert the approximate designs with values of possible continuous factors into exact designs with user-specified grid levels and the total number of experimental units.
By using this rounding algorithm, the yielded exact designs can maintain high relative efficiency with respect to the D-optimal approximate designs, and may further reduce the number of distinct experimental settings.

Following \cite{huang2024forlion} and \cite{lin2025ew}, the current {\bf ForLion} package concentrates on D-optimality, which is not only the most commonly used criterion in optimal design theory, but often leads to a design that performs well with respect to other criteria. Nevertheless, it can be extended to other criteria, given a corresponding lift-one algorithm being developed. A practical concern is its computational cost. As the numbers of factors and/or the levels of discrete factors increase, the overall optimization problem becomes more demanding. Although the current implementation is effective for a broad range of mixed-factor design problems, further improvements on computational efficiency in higher dimensional mixed-factor settings are important directions for future work.

\section*{Supplementary Material}

The Supplementary Material includes three sections: S1 provides quick reference tables for the arguments used in major {\bf ForLion} functions; S2 assesses the sensitivity of the locally D-optimal design against random seeds using the example in Section~\ref{sec:fly_example}; S3 extends the example in Section~\ref{sec:ESD_example} with a three-level discrete factor and an interaction term involving the continuous factor \texttt{Voltage}.

\clearpage
\renewcommand{\baselinestretch}{1}

%%%%%%%%%%%%%%%%%%%%%%%%%%%%%%%%%%%%%%%%%%%%%%%%%%%%%%%%%%%%%%%%%%%%%%%%%%%%%%%%%%%%%%%%%%%%%%%%%%%%%%%%%%%%%%%%%%%%%%%%%%%%

\centerline{\large\bf ForLion: An R Package for Finding Optimal Experimental Designs}
\vspace{2pt}
 \centerline{\large\bf with Mixed Factors}
%\vspace{2pt}
 %\centerline{\large\bf IF THIRD LINE IS NEEDED}
\vspace{.25cm}
 \centerline{Siting Lin$^{1}$, Yifei Huang$^{2}$, and Jie Yang$^{1}$} 
\vspace{.4cm}
 \centerline{\it $^1$University of Illinois at Chicago, $^2$Astellas Pharma Global Development, Inc.}
\vspace{.55cm}
 \centerline{\bf Supplementary Material}
\vspace{.55cm}
\fontsize{9}{11.5pt plus.8pt minus .6pt}\selectfont
\noindent

\normalsize

\noindent
{\bf S1 Function arguments:} We provide quick reference tables for the main functions in the {\bf ForLion} package; \\
{\bf S2 Assessing the sensitivity of the locally D-optimal design against random seeds:} We rerun the algorithm for the example in Section~\ref{sec:fly_example} with 10 randomly generated random seeds to assess the sensitivity of the obtained locally D-optimal design; \\
{\bf S3 Another example: A model with a three-level discrete factor and an interaction involving the continuous factor:} We consider a different model for the ESD experiment in Section~\ref{sec:ESD_example} involving a three-level discrete factor and an interaction term involving the continuous factor. 
\par

\setcounter{section}{0}
\setcounter{equation}{0}
\titleformat{\section}[hang]
  {\normalfont\large\bfseries}{\thesection}{1em}{}
\renewcommand\thesection{S\arabic{section}}

\setcounter{table}{0}
\renewcommand{\thetable}{S\arabic{table}}

\setcounter{figure}{0}
\renewcommand{\thefigure}{S\arabic{figure}}

\numberwithin{equation}{section}

%% set counter for supplementary material
\setcounter{page}{1}
\renewcommand{\thepage}{S.\arabic{page}}

%\fontsize{9}{12pt plus.8pt minus .6pt}\selectfont

\section{Function arguments}
\label{sec:function_arguments}

In this section, we provide a quick reference for the arguments used in the main functions of the {\bf ForLion} package. Table~\ref{tab:Special_arguments_for_the_ForLion_functions} lists the specialized arguments for \texttt{ForLion\_MLM\_Optimal} and \texttt{ForLion\_GLM\_Optimal}, which construct locally D-optimal approximate designs. Table~\ref{tab:Special_arguments_for_the_EW_ForLion_functions} summarizes the additional arguments for \texttt{EW\_ForLion\_MLM\_Optimal} and\\ \texttt{EW\_ForLion\_GLM\_Optimal}, which construct EW D-optimal approximate designs under unknown parameter values.
Table~\ref{tab:Comment_arguments} collects the arguments commonly used by the main functions.

\setcounter{table}{0}
\begin{table}[htbp]
\caption{Special arguments for two {\tt ForLion} functions}\label{tab:Special_arguments_for_the_ForLion_functions}
\centering
\footnotesize
\begin{tabularx}{\textwidth}{llL}  % 
\hline Function & Argument & Description \\
\hline 
\verb|ForLion_MLM_Optimal| & \verb|J| & Number of response categories in an MLM.\\
& \verb|bvec| & Vector of assumed parameter values.\\
%, same length as $h(y)$.\\
& \verb|Fi.func| & Function to calculate row-wise Fisher information Fi, default  \verb|Fi_MLM_func|.\\
& \verb|optim_grad| & TRUE or FALSE, default FALSE, whether  to use analytical or numerical gradient function when searching for a new design point.\\ \hline
\verb|ForLion_GLM_Optimal| & \verb|logscale| & TRUE or FALSE, whether or not to run the lift-one step in log-scale, i.e., using \verb|liftoneDoptimal_log_GLM_func()| or  \verb|liftoneDoptimal_GLM_func()|.\\
\hline
\end{tabularx}
\normalsize
\end{table}

\setcounter{table}{1}
\begin{table}[htbp]
\caption{Special arguments for two {\tt EW\_ForLion} functions}\label{tab:Special_arguments_for_the_EW_ForLion_functions}
\centering
\footnotesize
\begin{tabularx}{\textwidth}{llL}  % 
\hline Function & Argument & Description \\
\hline 
\verb|EW_ForLion_MLM_Optimal| 
& \verb|bvec_matrix| & Matrix of bootstrapped or simulated  parameter values.\\
& \verb|EW_Fi.func| & Function to calculate entry-wise expectation of Fisher information Fi, default  \verb|EW_Fi_MLM_func|.\\
&\verb|optim_grad|& TRUE or FALSE, default FALSE, whether  to use analytical or numerical gradient function when searching for a new design point.\\ \hline
\verb|EW_ForLion_GLM_Optimal| & \verb|joint_Func_b| & Prior distribution function of model parameters\\
& \verb|Integral_based|& TRUE or FALSE, whether or not integral-based EW D-optimality is used, FALSE indicates sample-based EW D-optimality is used.\\
& \verb|b_matrix|& Matrix of bootstrapped or simulated  parameter values.\\
&\verb|Lowerbounds|& Vector of lower ends of ranges of prior distribution for model parameters.\\
&\verb|Upperbounds|& Vector of upper ends of ranges of prior distribution for model parameters.\\
&\verb|logscale|& TRUE or FALSE, whether or not to run the lift-one step in log-scale, i.e., using \verb|EW_liftoneDoptimal_log_GLM_func()| or  \verb|EW_liftoneDoptimal_GLM_func()|.\\
\hline
\end{tabularx}
\normalsize
\end{table}

\setcounter{table}{2}
\begin{table}[htbp]
\caption{Common arguments in four key functions}\label{tab:Comment_arguments}
\centering
\footnotesize
\begin{tabularx}{\textwidth}{lL}  % 
\hline Argument & Description \\
\hline 
\verb|xlist_fix| & List of discrete-factor experimental settings under consideration, default NULL indicating a list of all possible discrete-factor experimental settings will be used.\\
\verb|h.prime| & User-supplied derivative function for continuous factors $\mathbf{x}_{(1)} \in \mathbb{R}^k$. 
For MLMs: return a list of $J\times p$ matrices $\partial \mathbf{X}_{\mathbf{x}}/\partial x_j$, $j=1,\ldots,k$ (numerical derivatives if omitted).  For GLMs: return the $p \times k$ matrix $\partial \mathbf{h}(\mathbf{x})/\partial \mathbf{x}_{(1)}$ (if omitted, support only main effects or interactions among discrete factors).\\
\verb|n.factor| & Vector of numbers of distinct levels, ``0'' indicating continuous factors that always come first, ``2'' or more for discrete factors, and ``1'' not allowed.\\
\verb|factor.level| & List of distinct factor levels, ``(min, max)'' for continuous factors that always come first, finite sets for discrete factors.\\
\verb|var_names| & Names for the design factors, Must have the same length as \verb|factor.level|; default ``X1'', ``X2'', ``X3'', etc. \\
\verb|hfunc| & Function for generating the corresponding model matrix or predictor vector, given an experimental setting or design point.\\
\verb|link| & Link function, for GLMs including ``logit'' (default), ``probit'', ``cloglog'', ``loglog'', ``cauchit'', ``log'', and ``identity'' (LM); for MLMs including ``continuation'' (default), ``baseline'', ``cumulative'', and ``adjacent''.\\
\verb|delta0|& Merging threshold for initial design, such that, $\| {\mathbf x}_i^{(0)} - {\mathbf x}_j^{(0)} \| \geq \delta_{0}$ for $i\neq j$, default $10^{-5}$.\\
\verb|epsilon| & Numerical tolerance. A nonnegative number is regarded as numerical zero if less than epsilon, default $10^{-12}$.\\
\verb|reltol| & Relative tolerance as converging threshold, default $10^{-5}$.\\
\verb|delta| & Relative difference as merging threshold for the merging step, two points satisfying $\|{\mathbf x}_i^{(t)} - {\mathbf x}_j^{(t)}\| < \delta$ may be merged, default $0$, can be different from {\tt delta0} for the initial design.\\
\verb|maxit| & Maximum number of iterations allowed, default $100$.\\
\verb|random|& TRUE or FALSE, whether or not to repeat the lift-one step multiple times with random initial allocations, 
default FALSE.\\
\verb|nram|& Number of times repeating the lift-one step with random initial allocations, valid only if {\tt random} is TRUE, default 3.\\
\verb|random.initial|& TRUE or FALSE, whether or not to repeat the whole procedure multiple times with random initial designs,  
default FALSE.\\
\verb|nram.initial|& Number of times repeating the whole procedure with random initial designs, valid only if {\tt random.initial} is TRUE, 
default 3.\\ 
\verb|rowmax| & Maximum number of points in initial designs, default NULL indicating no restriction.\\
\verb|Xini| & Initial list of design points, default NULL indicating automatically generating an initial list of design points.\\
\verb|pini| & A numeric vector specifying the initial weights for the design points in \verb|Xini|, default NULL indicating automatically generating an uniform weights of design points.\\
\hline
\end{tabularx}
\normalsize
\end{table}

\section{Assessing the sensitivity of the locally D-optimal design against random seeds}
\label{sec:assessing_sensitivity}

To assess the seed-to-seed variability of locally D-optimal designs for the example in Section~\ref{sec:fly_example}, we rerun the algorithm with 10 randomly generated random seeds (733, 633, 993, 294, 557, 623, 859, 108, 164 and 176), while keeping all other settings fixed.

For each run, we record the number of support points, the minimum distance among support points, and the relative D-efficiency with respect to the reference design obtained using \texttt{set.seed(123)} under the same algorithm settings.

\begin{figure}[ht]
  \centering
  \begin{subfigure}{0.6\textwidth}
    \centering
    \includegraphics[width=\textwidth]{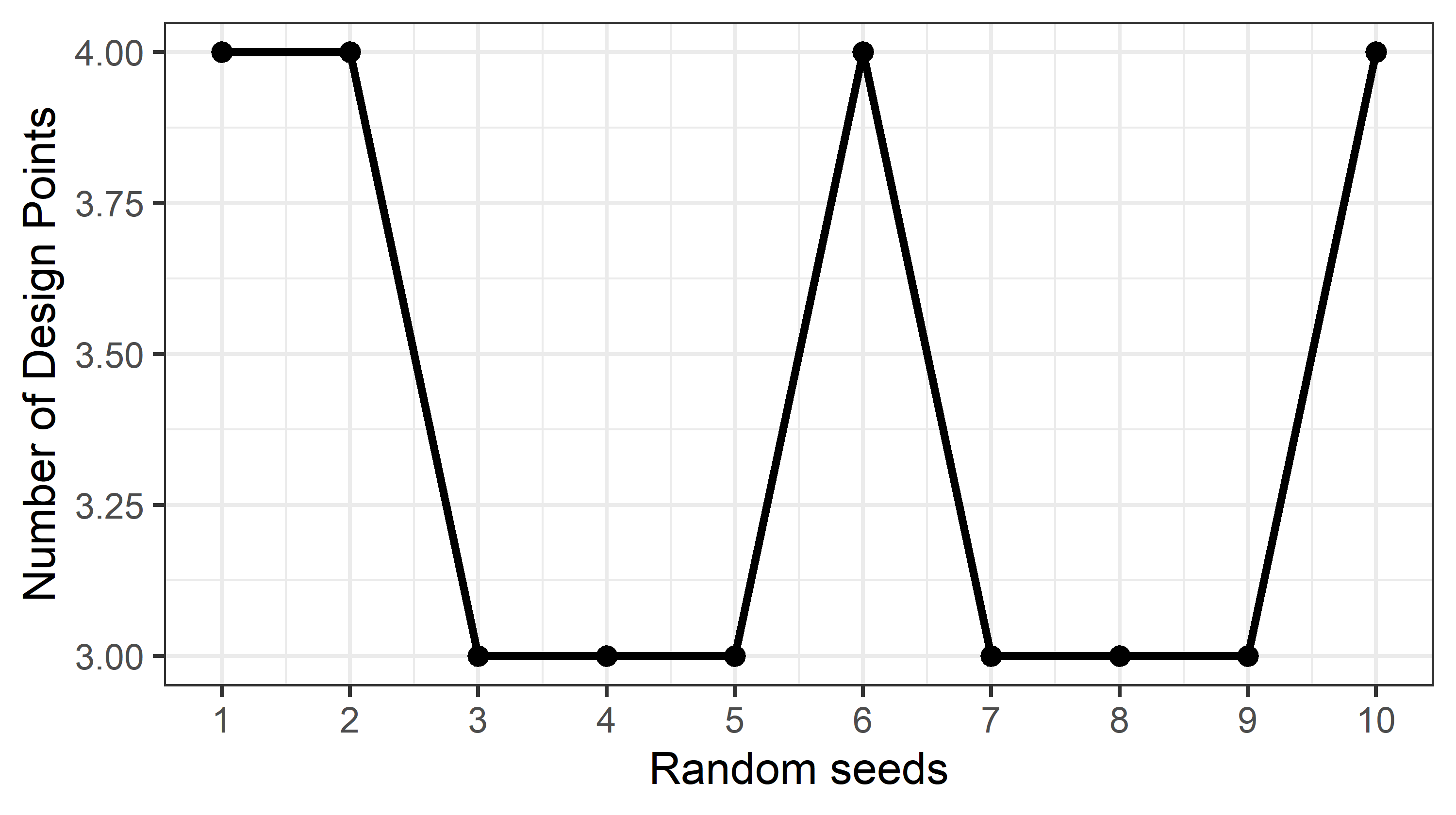}  
  \end{subfigure}
  \hfill
  \begin{subfigure}{0.7\textwidth}
    \centering
    \includegraphics[width=\textwidth]{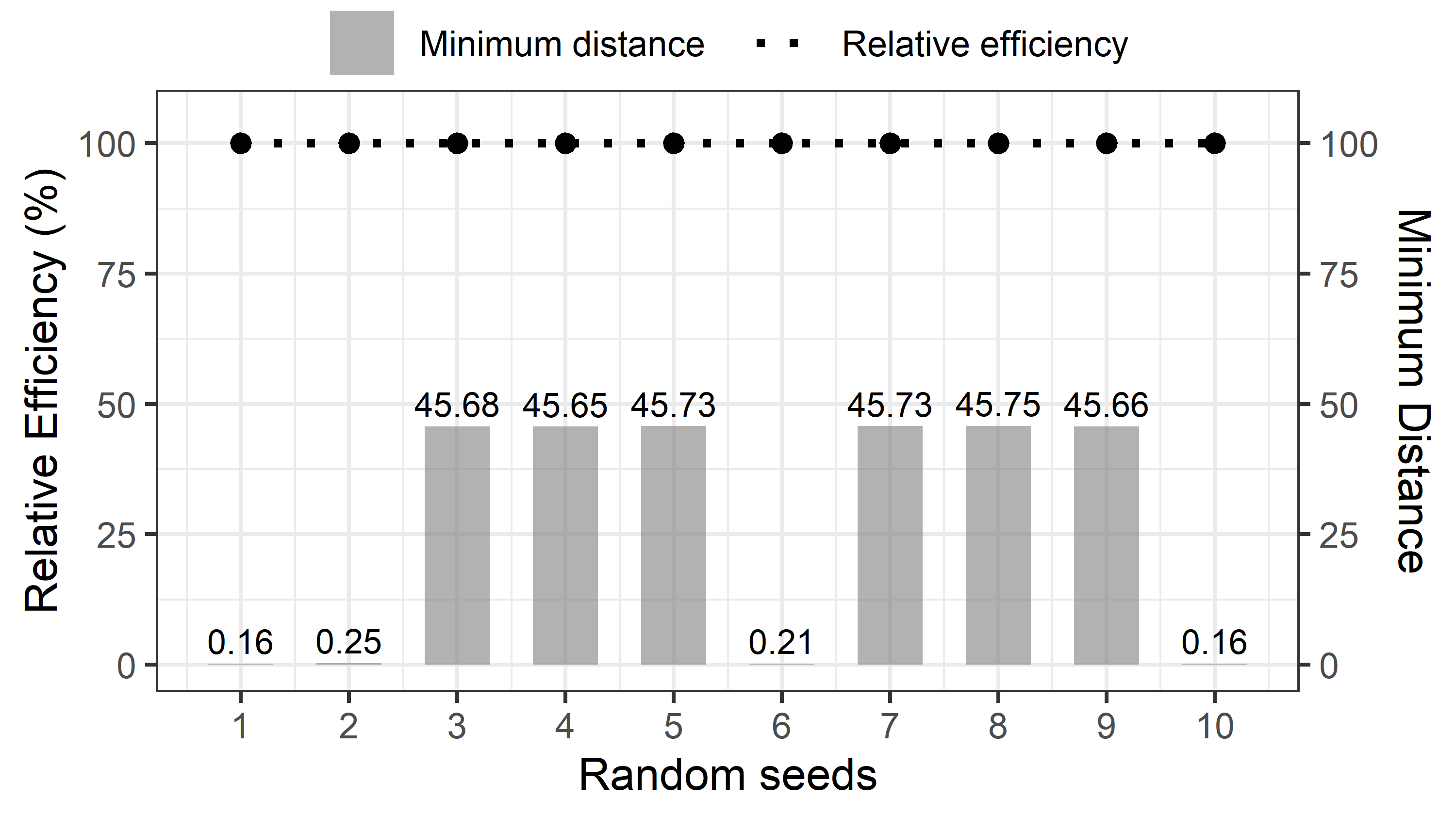} 
  \end{subfigure}\\ 
  \caption{Number of design points, minimum distance among design points (rounded to two decimals), and relative efficiency w.r.t. the design with $\delta=0.15$ and random seed 123}
  \label{fig:examp4_1_b}
\end{figure}

Figure~\ref{fig:examp4_1_b} summarizes the results. Across all runs, the relative efficiencies range from $99.99963\%$ to $100.0001\%$, which are essentially $100\%$ up to the specified numerical tolerance. The number of support points vary slightly across runs. These results indicate that different random seeds may lead to slightly different support points or weights, but the resulting D-efficiencies are practically the same in this example. 

\section{Another example: A model with a three-level discrete factor and an interaction involving the continuous factor}
\label{sec:additional_example_ESD}

For illustration purposes, we consider a different model for the ESD experiment in Section~\ref{sec:ESD_example} by treating \texttt{Pulse} as a three-level qualitative factor with levels $\{-1, \ 0, \ 1\}$ and adding an interaction between the continuous factor \texttt{Voltage} and \texttt{Pulse}. We use this example to demonstrate how a qualitative factor and its interaction with a continuous factor are implemented through \texttt{factor.level}, \texttt{hfunc}, and \texttt{h.prime}.

We retain the notation \texttt{LotA} ($x_1$), \texttt{LotB} ($x_2$), \texttt{ESD} ($x_3$), \texttt{Pulse} ($x_4$), and \texttt{Voltage}
($x_5$), where $x_5\in[25,45]$. We take $x_4=1$ as the reference level and define two dummy variables, namely
\[
{\tt Pul1}=\begin{cases}
    1, & \mbox{if }x_4=-1,\\
    0, & \text{otherwise;}
\end{cases}  \qquad  \text{and}  \qquad {\tt Pul2}= \begin{cases}
    1, & \mbox{if }x_4=0,\\
    0, & \text{otherwise.}
\end{cases}
\]
Then the three levels $-1,0,1$ of $x_4$ are represented by $({\tt Pul1},{\tt Pul2})=(1,0),(0,1),(0,0)$, respectively.

The resulting logistic regression model is
\[
 \operatorname{logit}(\mu)=\beta_0+\beta_1 x_1+\beta_2 x_2+\beta_3 x_3+\beta_{4} {\tt Pul1}+\beta_{5} {\tt Pul2}+\beta_6 x_5+\beta_{7}({\tt Pul1} \times x_5)+\beta_{8}({\tt Pul2} \times x_5)\ .  
\]
As in Section~\ref{sec:ESD_example}, we place the coefficient of the continuous factor $\beta_6$ first and the intercept last. We therefore reorder the model parameter vector and assume its nominal value as $\boldsymbol \theta = (\beta_6, \beta_1, \beta_2, \beta_3, \beta_4, \beta_5, \beta_{7}, \beta_{8}, \beta_0)^\top = (0.35, 1.50, -0.2, -0.15, 0.25, 0.40, 0.10, -0.05, -7.5)^\top$. Then the corresponding predictor vector becomes
\[
\mathbf{h}(\mathbf{x})=\bigl(x_5,\ x_1,\ x_2,\ x_3,\ {\tt Pul1},\ {\tt Pul2}, \ {\tt Pul1} \times x_5,\ {\tt Pul2} \times x_5,\ 1\bigr)^\top\ .
\]
We let $\mathbf{x}_{(1)}=x_5$ denote the continuous factor. Then the derivative used by the continuous search step is
\[
\frac{\partial \mathbf{h}(\mathbf{x})}{\partial \mathbf{x}^\top_{(1)}}=\bigl(1, \ 0, \ 0, \ 0, \ 0, \ 0, \ {\tt Pul1},\ {\tt Pul2}, \ 0\bigr)^\top\ .
\]
 As before, we reorder the factors as $(x_5,x_1,x_2,x_3,x_4)$. Therefore, the range of the continuous factor and the levels of the discrete factors are specified by
\[
\texttt{factor.level = list(c(25,45),c(-1,1),c(-1,1),c(-1,1),c(-1,0,1))}
\]

We then apply \texttt{ForLion\_GLM\_Optimal()} to obtain a locally D-optimal approximate design under this model. The corresponding R commands and results are listed below:
\footnotesize
\begin{verbatim}
## Note: Always put continuous factors ahead of discrete factors
## After reordering the components in x: x = (x5, x1, x2, x3, x4)
## Here x5--Voltage in [25,45], x1--LotA in {-1,1}, x2--LotB in {-1,1}, 
## x3--ESD in {-1,1}, x4--Pulse in {-1, 0, 1}
## Code x4 (x[5]) using two dummy variables: Pul1, Pul2
## x -> h(x) = (x5, x1, x2, x3, Pul1, Pul2, Pul1*x5, Pul2*x5, 1) 
>  hfunc.temp.int <- function(x) {
+                     Pul1 <- ifelse(x[5] == -1, 1, 0)
+                     Pul2 <- ifelse(x[5] == 0, 1, 0)  
+                     c(x[1:4], Pul1, Pul2, x[1]*Pul1, x[1]*Pul2, 1)}
# Adjust beta vector length accordingly 
#(now 9 terms: 4 main + 2 dummies + 2 interactions + intercept)
> beta.value.int <- c(0.35, 1.50, -0.2, -0.15, 0.25, 0.40, 0.10, -0.05, -7.5)
> variable_names.int <- c("Vol.", "LotA", "LotB", "ESD", "Pul.")
## hprime: derivative wrt x (dh(x)/d(x))
> hprime.temp.int <- function(x) {
+                    Pul1 <- ifelse(x[5] == -1, 1, 0)
+                    Pul2 <- ifelse(x[5] == 0, 1, 0)
+                    matrix(
+                      data = c(
+                        1,   # derivative wrt Vol.
+                        0,   # LotA
+                        0,   # LotB
+                        0,   # ESD
+                        0,   # Pul1
+                        0,   # Pul2
+                        Pul1,# interaction Vol*Pul1 wrt Vol
+                        Pul2,# interaction Vol*Pul2 wrt Vol
+                        0    # intercept
+                       ),
+                       nrow = 9, ncol = 1, byrow = TRUE)}
> set.seed(482)
> forlion_GLM_int <- ForLion_GLM_Optimal(n.factor = c(0, 2, 2, 2, 3),
+                    factor.level = list(c(25,45),c(-1,1),c(-1,1),c(-1,1),
+                    c(-1,0,1)), var_names = variable_names.int, 
+                    hfunc = hfunc.temp.int, h.prime = hprime.temp.int, 
+                    bvec = beta.value.int, link = "logit", delta0 = 0.01,
+                    epsilon = 1e-8, reltol = 1e-4, random = TRUE, 
+                    nram = 1, delta  = 0.08, maxit = 1200, logscale = TRUE)
\end{verbatim}
\begin{verbatim}
> print(forlion_GLM_int)
Design Output
============================================================== 
Count  Vol.     LotA     LotB     ESD      Pul.     Allocation
-------------------------------------------------------------- 
1      25.0000   1.0000  -1.0000   1.0000   0.0000  0.0308
2      25.0000   1.0000   1.0000  -1.0000   0.0000  0.0386
3      25.0000  -1.0000  -1.0000  -1.0000   1.0000  0.0869
4      25.0000  -1.0000  -1.0000   1.0000   1.0000  0.0141
5      25.0000  -1.0000   1.0000  -1.0000   1.0000  0.0134
6      32.8659  -1.0000  -1.0000   1.0000   0.0000  0.0490
7      34.3536  -1.0000   1.0000   1.0000   0.0000  0.0480
8      25.0000   1.0000   1.0000   1.0000   0.0000  0.0635
9      30.7937  -1.0000  -1.0000   1.0000   1.0000  0.0191
10     25.0000  -1.0000   1.0000   1.0000   1.0000  0.0870
11     31.1887  -1.0000   1.0000  -1.0000   1.0000  0.0280
12     25.0000  -1.0000  -1.0000   1.0000   0.0000  0.0673
13     33.2984  -1.0000   1.0000  -1.0000   0.0000  0.0527
14     32.2390  -1.0000   1.0000   1.0000   1.0000  0.0787
15     25.0000  -1.0000  -1.0000  -1.0000   0.0000  0.0304
16     25.0000  -1.0000   1.0000  -1.0000  -1.0000  0.0110
17     29.6334  -1.0000   1.0000   1.0000  -1.0000  0.1111
18     25.0014  -1.0000   1.0000   1.0000  -1.0000  0.1027
19     25.0602  -1.0000   1.0000  -1.0000   0.0000  0.0679
============================================================== 
m:
[1] 19
det:
[1] 6.404087e-10
convergence:
[1] TRUE
min.diff:
[1] 1
x.close:
     [,1] [,2] [,3] [,4] [,5]
[1,]   25   -1   -1   -1    1
[2,]   25   -1   -1   -1    0
itmax:
[1] 1160
\end{verbatim}
\normalsize

\end{document}